\documentclass[aps,floatfix,eqsecnum,showpacs,preprint,tightenlines]{revtex4-1}
\usepackage{epsfig}
\usepackage{bm}

\usepackage{amsmath}

\def\e{\kern+.5ex\lower.42ex\hbox{$\scriptstyle \iota$}\kern-1.10ex e}
\def\registered{{\ooalign{\hfil\raise .00ex\hbox{\scriptsize R}\hfil\crcr\mathhexbox20D}}}

\newcommand{\BA}[1]{\langle #1 \mid}

\newcommand{\KT}[1]{\mid #1 \rangle}

\begin{document}


\title{Radiative pion capture in $^2$H, $^3$He and $^3$H}



\author{J. Golak}
\affiliation{M. Smoluchowski Institute of Physics, Jagiellonian University, PL-30348 Krak\'ow, Poland}
\author{R. Skibi{\'n}ski}
\affiliation{M. Smoluchowski Institute of Physics, Jagiellonian University, PL-30348 Krak\'ow, Poland}
\author{K. Topolnicki}
\affiliation{M. Smoluchowski Institute of Physics, Jagiellonian University, PL-30348 Krak\'ow, Poland}
\author{H. Wita{\l}a}
\affiliation{M. Smoluchowski Institute of Physics, Jagiellonian University, PL-30348 Krak\'ow, Poland}
\author{A. Grassi}
\affiliation{M. Smoluchowski Institute of Physics, Jagiellonian University, PL-30348 Krak\'ow, Poland}
\author{H. Kamada}
\affiliation{Department of Physics, Faculty of Engineering,
Kyushu Institute of Technology, Kitakyushu 804-8550, Japan}
\author{A. Nogga}
\affiliation{Institute for Advanced Simulation, Institut f\"ur Kernphysik, J\"ulich Center for Hadron Physics, and 
             JARA - High Performance Computing,  Forschungszentrum J\"ulich, D-52425 J\"ulich, Germany}
\author{L. E. Marcucci}
\affiliation{Department of Physics, University of Pisa, IT-56127 Pisa, Italy
              and INFN-Pisa, IT-56127 Pisa, Italy}


\date{\today}

\begin{abstract}
The
$\pi^- + {^2{\rm H}} \rightarrow \gamma + n + n$, 
$\pi^- + {^3{\rm He}} \rightarrow \gamma + {^3{\rm H}}$, 
$\pi^- + {^3{\rm He}} \rightarrow \gamma + n + d$, 
$\pi^- + {^3{\rm He}} \rightarrow \gamma + n + n + p$
and 
$\pi^- + {^3{\rm H}} \rightarrow \gamma + n + n + n$
capture reactions
are studied with the AV18 two-nucleon potential and the Urbana~IX
three-nucleon potential. We provide for the first time 
realistic predictions 
for the differential and total capture rates for all these processes,
treating consistently the initial and final nuclear states.
Our results are based on the single nucleon Kroll-Ruderman-type transition operator
and concentrate on the full treatment of the nuclear final state 
interactions. 
They are compared with older theoretical predictions 
and experimental data.
\end{abstract}

\pacs{23.40.-s, 21.45.-v, 27.10.+h}

\maketitle


\section{Introduction}
\label{section1}

Studies of radiative pion capture were initiated in 1951
by Panofsky, Aamodt and Hadley \cite{Panofsky1951}, who measured
the ratio of mesonic ($\pi^- + p \rightarrow n + \pi^0$) 
to radiative ($\pi^- + p \rightarrow n + \gamma$) 
capture of stopped negative pions in hydrogen. 
Measurements of capture reactions in the early 1950s played an important role in 
fixing fundamental properties of the pions and their interactions 
with the nucleons. Later, experiments performed at the Lawrence 
Radiation Laboratory in Berkeley delivered photon spectra 
from radiative pion capture on different nuclei. Many such measurements
with improved resolution were conducted in the 1970s 
at the Swiss Institute for Nuclear Research (SIN) 
(later the Paul Scherrer Institute (PSI))
by a collaboration including the Universities of Lausanne, Munich and Zurich.
All the early experimental and theoretical 
work prior to January 1976 was summarized in Ref.~\cite{Baer1977}.
Since we restrict ourselves to reactions 
with two and three nucleons in the present paper, we only refer the reader to 
Refs.~\cite{Renker1978,Reynaud1981,Martoff1983,Baer1983,Roig1985,Singham1986,Krivine1988,Navarro1989,Raywood1997,Amaro1997}
for later studies of systems with $A>3$. 

The studies of the $\pi^- + {^2{\rm H}} \rightarrow \gamma + n + n$ reaction 
concentrated on the extraction of the $^1S_0$ neutron-neutron scattering length, $a_{nn}$.
This reaction produces three detectable particles in a final state 
and the interaction of the photon with two emerging neutrons 
is so weak that the final state interaction (FSI) is absolutely dominated
by the neutron-neutron (nn) scattering. 
As early as in 1951 Watson and Stuart \cite{Watson1951} showed
with quite simple dynamics that the corresponding photon spectrum very strongly depends 
 on the properties of the nn interaction.
Since then many theoretical efforts
\cite{McVoy1961,Bander1964,Gibbs1975,Gibbs1977,deTeramond1977,deTeramond1987,Gardestig2006PRC1,Gardestig2006PRL}
combined with more and more precise
measurements 
\cite{Phillips1954,Ryan1964,Haddock1965,Nicholson1968,Salter1975,Gabioud1979,Gabioud1981,Gabioud1984,Schori1987,Howell1998,Chen2008},
have contributed decisively to our present day knowledge about $a_{nn}$. 
Detailed information about this rich field can be found in the review by \v{S}laus, Akaishi, and Tanaka~\cite{Slaus1989}
and in the more recent review by G{\aa}rdestig \cite{Gardestig2009}, 
where also complementary efforts \cite{Huhn2000,GonzalezTrotter2006} to determine $a_{nn}$ from
the ${^2{\rm H}}(n,np)n$ reaction were reported.

In order to extract $a_{nn}$ from the $\pi^- + {^2{\rm H}} \rightarrow \gamma + n + n$ process,
various theoretical frameworks were employed.
The approach formulated and applied by Gibbs, Gibson, 
and Stephenson~\cite{Gibbs1975,Gibbs1977.1,Gibbs1977}
used a nonrelativistic one-body transition operator
containing relativistic corrections and a nn
scattering wave function generated from the Reid soft-core
potential~\cite{Reid1968}. The wave function was obtained
in coordinate space, starting from the asymptotic
region and then integrating towards smaller inter-nucleon distances $r$.
For $ r\le 1.4$~fm a fifth degree polynomial with appropriate boundary 
conditions was used to represent the wave function. 
In order to minimize the error in the $a_{nn}$ extraction, the calculations
had to be restricted to small relative nn
energies, that is to the nn FSI peak region. While the first analysis 
of the experiment conducted 
at the Clinton P. Anderson
Meson Physics Facility at Los Alamos (LAMPF)
reported 
in Ref.~\cite{Howell1998} used theoretical cross sections derived with 
nonrelativistic phase space factors, the results published 
later in Ref.~\cite{Chen2008} were obtained with the corresponding 
relativistic formulas.

The second theoretical approach
treated the nn rescatterings
by means of Muskhelishvili-Omn\`es dispersion 
relations~\cite{Muskhelishvili,Omnes}.
De T\'eramond and collaborators~\cite{deTeramond1977,deTeramond1980,deTeramond1987} considered various 
dynamical ingredients, like pion rescattering terms, off-shell
effects, the impact of higher partial waves and studied their importance 
for the extraction of $a_{nn}$. This theoretical framework 
was employed in the analysis of two experiments performed 
at SIN~\cite{Gabioud1979,Gabioud1981,Gabioud1984,Schori1987}. 
It is quite remarkable that the analyses of the SIN 
and LAMPF experiments led to equivalent results for $a_{nn}$,
with very small theoretical errors of 0.3~fm in $a_{nn}$,
even though they were based on different theories.
Namely, the final result from the SIN experiment~\cite{Schori1987}
was $a_{nn} = (-18.7 \pm 0.6)$~fm, representing a weighted mean of two data sets with
systematic and theoretical errors added in quadrature.
The corresponding result from the LAMPF experiment reported
in Ref.~\cite{Chen2008} read $a_{nn} = (-18.63 \pm 0.27 \, {\rm (experiment)} \pm 0.30 \, {\rm (theory)} )$~fm.

Further progress in the theoretical treatment of radiative pion capture 
reactions and the inverse process, pion photoproduction, 
was made in the framework of chiral effective field theory. 
In particular neutral pion photoproduction from a nucleon 
was studied by Bernard and collaborators \cite{Bernard1996.1} 
in the framework of heavy baryon chiral perturbation
theory (HBCHPT). The same authors calculated also 
the one-loop corrections to the Kroll-Ruderman low-energy 
theorems for charged pion photoproduction at threshold
in Ref.~\cite{Bernard1996.2}.
Within the same framework Fearing {\em et al.}~\cite{Fearing2000}
evaluated the transition amplitude for the 
photoproduction process away from threshold 
and obtained expressions for the $s$- and $p$-wave multipoles.
They made connection with the radiative capture
reaction at the cross section level via the detailed balance equation.

Several years later, G{\aa}rdestig and Phillips
applied HBCHPT to the $\pi^- + {^2{\rm H}} \rightarrow \gamma+n+n$ 
reaction~\cite{Gardestig2006PRC1,Gardestig2006PRC2,Gardestig2006PRL} and for the first time used a consistent 
transition operator (with one- and two-body contributions)
as well as the deuteron and nn 
scattering states. Namely, they worked in coordinate space
and, starting from the asymptotic state, calculated the 
nn wave function solving the Schr\"odinger equation, which 
contained the lowest order chiral potential.
For distances $r$ smaller than a few fermis, a solution for the spherical
well potential was chosen to account for the unknown short-distance physics.
First calculations of G{\aa}rdestig and Phillips were carried out 
at next-to-next-to-leading order of chiral expansion
and made it possible to extract $a_{nn}$ with a precision at the $0.2$~fm level.

Further investigations \cite{Gardestig2006PRC2,Gardestig2006PRL} at higher order of chiral expansion
revealed important relations between the short-distance
physics in a number of reactions on light nuclei, since 
the same axial isovector two-body contact term was found to contribute 
in the radiative pion capture on the deuteron, pion production in nucleon-nucleon (NN) scattering, 
triton $\beta$ decay, proton-proton fusion, neutrino-deuteron scattering, 
muon capture on the deuteron, nucleon-deuteron scattering and the 
$p + {^3{\rm He}} \rightarrow  {^4{\rm He}} + e^+ + \nu_e $  (hep) process.
The correlation found by G{\aa}rdestig and Phillips
had direct impact on the accuracy with which $a_{nn}$ was estimated,
and allowed them to reduce the theoretical error to approximately $0.05$~fm.

Investigation of the radiative (and non-radiative) negative pion capture 
in the three-nucleon (3N) bound states also started in the early 1950s. Even if this 
work prior to January 1976 is described in Ref.~\cite{Baer1977}, we mention 
here a pioneering contribution by Messiah~\cite{Messiah1952}, who 
formulated a theoretical framework to deal with pion capture (not only radiative),
discussed various dynamical aspects and gave predictions for all 
the six reaction channels for the $^3$He nucleus, which were 
however based on very crude approximations in the closure formulas.

Further theoretical efforts to describe radiative capture concentrated 
on the $\pi^- + {^3{\rm He}} \rightarrow \gamma + {^3{\rm H}}$ process.
Some authors used very simple parametrizations of the $^3$He and $^3$H 
wave functions 
and the Kroll-Ruderman~\cite{KR1954} form of the transition operator 
to calculate directly the pertinent nuclear matrix elements
(see for example Ref.~\cite{Divakaran1965}).
In other papers the so-called ``elementary particle treatment of nuclei'' 
was adopted~\cite{Ericson1967,Ericson1969,Griffiths1968,Pascual1970}. Since the early 
measurements~\cite{Zaimidoroga65,Zaimidoroga67} yielded no absolute capture 
rates but rather relative probabilities of various processes,
the theories tried to reproduce measured branching ratios. In particular
the so-called Panofsky ratio, that is the ratio
of the probabilities 
of the charge exchange 
$\pi^- + {^3{\rm He}} \rightarrow \pi^0 + {^3{\rm H}}$ 
and radiative capture 
$\pi^- + {^3{\rm He}} \rightarrow \gamma + {^3{\rm H}}$ 
reactions was studied in many papers, as exemplified 
by Refs.~\cite{Phillips1974,Gibbs1978}.
An important paper by Tru\"ol {\em et al.}~\cite{Truoel1974} not only brought 
new experimental data on the photon spectrum and several branching ratios 
but also corrections to the earlier theoretical predictions 
published in Refs.~\cite{Fuji1962,Divakaran1965,Pascual1970}.
Using a formula from Ref.~\cite{Delorme1966} and making 
a connection between the Gamow-Teller 
matrix element in triton $\beta$-decay and the corresponding 
matrix elements in non-breakup pion radiative capture on $^3$He, the authors 
of Ref.~\cite{Truoel1974} also provided  a result for the 
$\pi^- + {^3{\rm He}} \rightarrow \gamma + {^3{\rm H}}$ capture rate.

The photon spectrum and the branching ratios from 
Ref.~\cite{Truoel1974} were analyzed by 
Phillips and Roig~\cite{Phillips1974}. Radiative breakup rates were 
calculated in the impulse approximation and
FSI among the three nucleons 
was treated in the Amado model~\cite{Amado1963,Lovelace1964}, 
by solving the Faddeev equations with a simple separable $s$-wave 
NN potential. The 3N bound states 
were not calculated consistently but had an analytical form, 
which allowed the authors to regulate strengths
of the principal $S$-state, $S^\prime$-state and $D$-state 
components. Despite these simplifications, Phillips and Roig 
could describe the shapes of the experimental photon spectrum 
and the branching ratios given in Ref.~\cite{Truoel1974}. 
They also predicted the decisive role of FSI
and the dominant contribution of the 
$\pi^- + {^3{\rm He}} \rightarrow \gamma + n + d$ channel in the radiative
breakup of $^3$He.

In Ref.~\cite{Phillips1975}, the same authors made calculations
for the 
$\pi^- + {^3{\rm H}} \rightarrow \gamma + n + n + n$ capture reaction, 
later confronted with 
experimental results obtained by Bistirlich {\em et al.} \cite{Bistirlich1976}
and (in an improved experiment) by Miller {\em et al.} \cite{Miller1980}.
This reaction allows one to study the 3N 
system in a pure total isospin $T= 3/2$ state and to search for resonant or 
even bound three-neutron states. None were found in the two above-mentioned
LAMPF experiments and the smooth shape of the experimental photon 
spectrum was in satisfactory 
agreement with the theoretical predictions by Phillips and Roig.
In particular FSI raised the theoretical spectrum
obtained under a plane wave impulse approximation. 
Since later all pion beam facilities were shut down and the more recent theoretical work 
on radiative pion capture was focused on the single-nucleon and two-nucleon (2N) sector, 
no calculations for the 3N system with modern realistic nuclear forces have been
performed.

Recently we have established a theoretical framework for
the $A\leq 3$ muon capture reactions~\cite{PRC90.024001,PRC94.034002}. Important building blocks 
of this framework were cross-checked with the results 
from Ref.~\cite{prc83.014002}, obtained using the hyperspherical harmonics
formalism~\cite{Kie08}.
It has then become  very natural to adapt our momentum space techniques
for corresponding radiative capture reactions. 
Thus we provide, for the first time, predictions 
with consistent treatment of the initial and final nuclear states 
calculated from realistic 2N and 3N forces
for the differential and total capture rates of the
$\pi^- + {^2{\rm H}} \rightarrow \gamma + n + n$,
$\pi^- + {^3{\rm He}}  \rightarrow \gamma + {^3{\rm H}}$,
$\pi^- + {^3{\rm He}} \rightarrow \gamma + n + d$,
$\pi^- + {^3{\rm He}} \rightarrow \gamma + n + n + p$,
and 
$\pi^- + {^3{\rm H}} \rightarrow \gamma + n + n + n$ reactions.

The paper is organized in the following way.
In Sec.~\ref{section2} we introduce the single nucleon 
transition operator. 
In the following sections, we show selected results, concentrating on the 
photon spectra and total capture rates. In Sec.~\ref{section3}, we start with the 
$\pi^- + {^2{\rm H}} \rightarrow \gamma + n + n$
reaction and demonstrate that our framework
possesses the same sensitivity to $a_{nn}$ as the older \cite{Schori1987,Chen2008} 
and more recent, chiral calculations \cite{Gardestig2006PRC1}.
Our results concerning 
$\pi^- + {^3{\rm He}} \rightarrow \gamma + {^3{\rm H}}$
and breakup reactions with trinucleons 
are described in further sections, where we compare 
predictions obtained with different treatment of FSI.
All our results for the total capture rates are shown together and compared to earlier theoretical predictions
in Sec.~\ref{section7}.
Finally, Sec.~\ref{section8} contains our summary and outlook.

\section{The transition operator}
\label{section2}

The radiative capture process is treated in the same way as muon capture
from the $1s$ atomic orbit.
Namely, the initial state 
$ \KT{i\, } $ comprises the $K$-shell pion wave function 
$ \KT{ \psi \, } $ 
and the initial nucleus state 
$\KT{\Psi_{\!i} \, {\bf P}_{\!i} \, m_{i} \, } $
with the three-momentum ${\bf P}_{\!i}$ 
and the spin
projection $m_{i}$:
\begin{eqnarray}
\KT{i\, } = \KT{\psi } \, \KT{\Psi_{\!i} \, {\bf P}_{\!i} \, m_{i} \, } \, .
\label{i}
\end{eqnarray}
In the final state $\KT{f\, }$
the photon occurs,
described by the state 
$\KT{\gamma \, {\bf p}_{\gamma} \, {\bm \epsilon} \, } $
with the three-momentum ${\bf p}_{\gamma}$ 
and the polarization vector ${\bm \epsilon}$ perpendicular to ${\bf p}_{\gamma}$, 
accompanied by the final nuclear state 
$\KT{\Psi_{\!f} \, {\bf P}_{\!f} \, m_{f} \, } $
with the total 
three-momentum ${\bf P}_{\!f}$ and the set of spin projections $m_{f}$:
\begin{eqnarray}
\KT{f\, } = \KT{\gamma \, {\bf p}_{\gamma} \, {\bm \epsilon} \, } \, 
\KT{\Psi_{\!f} \, {\bf P}_{\!f} \, m_{f} \, } \, .
\label{f}
\end{eqnarray}
We assume that the transition from the initial to final state is governed by the 
one-body Kroll-Rudermann operator $j_{KR}$ \cite{KR1954} and start with 
the simple nonrelativistic form from Ref.~\cite{Baer1977}. It is given as 
\begin{eqnarray}
j_{KR} = -i e \,  \frac {g_{A} } { g_{V} } \, \frac1{ f_\pi } {\bm \epsilon} \cdot {\bm \sigma} \, {\tau}_{-} \, ,
\label{jKR}
\end{eqnarray}
where $e$, $g_A$, $g_V$, $f_\pi$ and ${\bm \sigma}$ are 
the elementary charge, axial-vector and vector coupling constants,
the pion decay constant and the nucleon spin operator, respectively. 
In the isospin formalism, also the isospin lowering operator,
${\tau}_{-}=(\tau_x -{\rm i} \tau_y)/2$, is introduced.
To make a simple connection to our work on muon capture \cite{PRC90.024001,PRC94.034002}
we replace Eq.~(\ref{jKR}) by
\begin{eqnarray}
j_{KR} = -i \, \frac{e}{ f_\pi } {\bm \epsilon} \cdot {\bf j}_A \, ,
\label{jKR2}
\end{eqnarray}
where ${\bf j}_A$ is the single nucleon axial current from Refs.~\cite{PRC86.035503,PRC90.024001}.
Its matrix elements in momentum space of one nucleon read
\begin{eqnarray}
&&\BA{{\bf p}^{\, \prime}} {{\bf j}}_{\text{A}}(1)  \KT{{\bf p}} = 
\bigg\{ g_{1}^{A} \left(1 - \frac{\left({\bf p} + {\bf p}^{\, \prime} \, \right)^{2}}{8 M^{2}} \, \right) {{\bm \sigma}} 
\nonumber \\
&& + \, \frac{g_{1}^{A}}{4 M^{2}} \big[ \left({\bf p} \cdot {{\bm \sigma}} \, \right) {\bf p}^{\, \prime} + \left({\bf p}^{\, \prime} 
\cdot {{\bm \sigma}} \, \right) {\bf p} + \, i \, \left( {\bf p} \times 
{\bf p}^{\, \prime} \,  \right) \big]  \nonumber \\
&& + \, g_{2}^{A} \left( {\bf p} - {\bf p}^{\, \prime} \, \right) \frac{{{\bm \sigma}} \cdot \left({\bf p} - {\bf p}^{\, \prime} \, \right)}{2 M} 
\bigg\} {\tau}_{-} \, ,
\label{jA1}
\end{eqnarray}
where $M \approx 939$~MeV is the nucleon mass 
and the detailed information about the nucleon form factors 
$g_{1}^{A}$ and $g_{2}^{A}$ is provided in Ref.~\cite{PRC86.035503}.

In practice it is always possible to set $ {\hat {\bf p}}_\gamma = - {\hat {\bf z}} $,
which means that for a real photon only 
two nuclear matrix elements of the transverse components of ${\bf j}_A$ (here represented in the 
spherical notation) need to be calculated
\begin{eqnarray}
{N}_{\pm 1} = 
\BA{\Psi_{\!f} \, {\bf P}_{\!f} \, m_{f} \, } \, 
j_{A, \pm 1} 
\, \KT{\Psi_{\!i} \, {\bf P}_{\!i} \, m_{i} \, } \, .
\label{nlambda}
\end{eqnarray}
The transversality condition implies also that the term in Eq.~(\ref{jA1}) proportional 
to $g_2^A$ does not contribute.

The form of the transition operator employed in this article definitely leaves 
room for improvement. Radiative pion capture by a single nucleon 
and the inverse, photoproduction, process was studied in heavy baryon chiral perturbation 
theory in Refs.~\cite{Bernard1996.1,Bernard1996.2,Fearing2000} 
and corrections to the Kroll-Ruderman
low-energy theorem were calculated. 
One can also expect many-nucleon, most importantly 2N, contributions to the capture
process. They were derived more recently by G{\aa}rdestig \cite{Gardestig2006PRC1,Gardestig2006PRC2,Gardestig2006PRL}.

Our first predictions, however, are based on the single nucleon transition 
operator and focus on other dynamical ingredients. 
Like for muon capture, we want to concentrate on FSI
in the nuclear sector. To this end we calculate 2N and 3N scattering 
states using the AV18 NN potential \cite{av18} and the Urbana~IX 3N force \cite{urbana}.
To the best of our knowledge, we provide, for the first time, 
consistent predictions for the total capture rates of the 
$\pi^- + {^2{\rm H}} \rightarrow \gamma + n + n$,
$\pi^- + {^3{\rm He}}  \rightarrow \gamma + {^3{\rm H}}$,
$\pi^- + {^3{\rm He}} \rightarrow \gamma + n + d$,
$\pi^- + {^3{\rm He}} \rightarrow \gamma + n + n + p$,
and 
$\pi^- + {^3{\rm H}} \rightarrow \gamma + n + n + n$ reactions,
obtained with realistic 2N and 3N forces.

Many elements of our calculations for all the listed reactions 
are essentially the same as performed for the corresponding 
muon capture reactions in Refs.~\cite{PRC90.024001,PRC94.034002}.
In particular the formulas concerning kinematics can be directly used,
if the muon mass is replaced by the negative pion mass.
The radiative pion capture rates for the totally unpolarized reactions 
are also easily obtained from the corresponding 
expressions for the muon capture rates. 

\section{Results for the $\pi^- + {^2{\rm H}} \rightarrow \gamma + n + n $ reaction}
\label{section3}

Our description of nuclear initial and final states is based on the nonrelativistic potentials 
and dynamical equations. It should then be  used with the 
nonrelativistic kinematics. It is then mandatory to verify 
if the nonrelativistic approximations in the kinematics of the 
nuclear sector (the photon is of course treated relativistically)
is justified. Clearly the pion is heavier than the muon, so pion
absorption brings more energy to the nuclear system. Thus the comparisons 
of various results computed from the 
nonrelativistic and relativistic nuclear kinematics 
performed in Refs.~\cite{PRC90.024001,PRC94.034002} for muon capture 
had to be repeated with a new mass of the absorbed particle.

Starting from the energy and momentum conservation,
we obtain first the maximal relativistic and nonrelativistic 
photon energies:
\begin{eqnarray} 
\left( E_\gamma^{max,nn} \, \right)^{rel} =
\frac{1}{2} \left(-\frac{4 {M_n}^2}{{M_d}+{M_\pi}}+ {M_d}+{M_\pi}\right)
\end{eqnarray}  
and
\begin{eqnarray} 
\left( E_\gamma^{max,nn} \, \right)^{nrl} =
2 \sqrt{{M_d} {M_n}+{M_\pi}
   {M_n}-{M_n}^2}-2 {M_n} \, .
\end{eqnarray}  
Assuming 
$ M_p$ = 938.272 MeV,
$ M_n$ = 939.565 MeV,
$M_\pi$ = 139.570 MeV,
$M_d = M_p + M_n$ - 2.225 MeV, we obtain
$ \left( E_\gamma^{max,nn} \, \right)^{rel} $ = 131.459 MeV
and
$ \left( E_\gamma^{max,nn} \, \right)^{nrl} $ = 131.454 MeV, respectively,
with a difference which is clearly negligible.

\begin{figure}
\includegraphics[width=8cm]{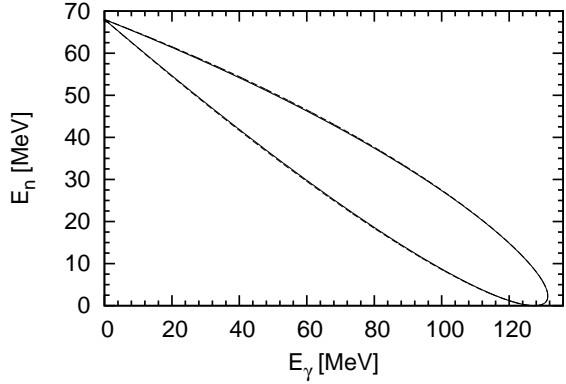}
\caption{The kinematically allowed region in the
$E_\gamma - E_n$ plane calculated relativistically
(solid curve) and
nonrelativistically (dashed curve) for the
$\pi^-+ {^2{\rm H}} \rightarrow \gamma + n + n $ capture process. 
The lines practically overlap.
\label{egamma_2H_en}}
\end{figure}

In Fig.~\ref{egamma_2H_en}, we demonstrate additionally 
that the kinematically allowed regions in the
$E_\gamma - E_n$ plane calculated relativistically
and nonrelativistically essentially overlap, which means 
that the nonrelativistic kinematics can be safely used.

Taking the form of the transition matrix element into account, introducing the
bosonic factors from the pion and photon fields
and evaluating the phase space factor 
in terms of the relative nn momentum, 
$ {\bf p} = \frac12 \,  \left( {\bf p}_1 -  {\bf p}_2 \, \right) $
(${\bf p}_1$ and ${\bf p}_2$ are the two individual neutron momenta), 
we arrive at the following 
expression for the total unpolarized capture rate
\begin{eqnarray}
&&\Gamma_{nn} = \frac12 \frac1{( 2 \pi )^2 } \, 
\frac{ 2 \pi \alpha }{f_\pi^2 M_\pi } \,
\frac { \left(  M^\prime_d \alpha \, \right)^3 } {\pi  } \, 
\int\limits_{0}^{\pi} d \theta_{p_\gamma}  \sin \theta_{p_\gamma}  \, \int\limits_{0}^{2 \pi} d \phi_{p_\gamma}  \, 
\int\limits_0^{E_\gamma^{max,nn}} \, dE_\gamma E_\gamma  \,
\frac12 M_n p  \,  
\nonumber \\
&&\int\limits_{0}^{\pi} d \theta_p \sin \theta_p \, \int\limits_{0}^{2 \pi} d \phi_p \, 
\frac13 \sum\limits_{m_d} \sum\limits_{m_1, m_2} 
\Big( 
\left| N_{+1}(m_1, m_2, m_d \, ) \, \right|^2  
\, + \, 
\left| N_{-1}(m_1, m_2, m_d \, ) \, \right|^2  \, 
\Big) \, ,
\label{gnn1}
\end{eqnarray}  
where 
the factor $ \frac { \left(  M^\prime_d \alpha \, \right)^3 } {\pi  } $ stems from the 
$K$-shell atomic wave function, $ M^\prime_d  = \frac { M_d M_\pi } { M_d + M_\pi }$,
$ \alpha \approx \frac 1{137} $ is the fine structure constant 
and $p \equiv \left| {\bf p} \right| = p ( E_\gamma\, ) $. 
We use in our calculations $ f_\pi \equiv \sqrt{2} F_\pi = 0.932\, M_\pi$~\cite{Baer1977}.

We can further simplify Eq.~(\ref{gnn1}), since for the unpolarized case only 
the relative angle between ${\bf p}$ and ${\bf p}_\gamma$ matters.
Therefore we set $ {\hat {\bf p}}_\gamma = - {\hat {\bf z}} $ and
choose the azimuthal angle of the relative momentum $\phi_p=0$,
which leads to:
\begin{eqnarray}
&&\Gamma_{nn} = \frac12 \frac1{( 2 \pi )^2 } \, 
\frac{ 2 \pi \alpha }{f_\pi^2 M_\pi } \,
\frac { \left(  M^\prime_d \alpha \, \right)^3 } {\pi  } \, 
4 \pi \,
\int\limits_0^{E_\gamma^{max,nn}} \, dE_\gamma E_\gamma  \,
\frac12 M_n p  \, 2 \pi \,  
\nonumber \\
&& \int\limits_{0}^{\pi} d \theta_p \sin \theta_p \, 
\frac13 \sum\limits_{m_d} \sum\limits_{m_1, m_2} 
\Big( 
\left| N_{+1}(m_1, m_2, m_d \, ) \, \right |^2 \, + \, 
\left| N_{-1}(m_1, m_2, m_d \, ) \, \right |^2 \, 
\Big) \, .
\label{gnn2}
\end{eqnarray}  

We generate the nuclear matrix elements 
$N_{\pm 1}(m_1, m_2, m_d \, ) $ in momentum space \cite{PRC90.024001}.
They contain plane wave as well as rescattering contributions.
Although the three-dimensional formalism of Ref.~\cite{edis3d}
could be applied also to radiative pion capture in $^2$H, in this article 
we discuss results obtained solely with a standard partial wave decomposition (PWD). 
In the calculations, we
exclusively used  the AV18 NN potential~\cite{av18}. 
However, based on our experience from the muon capture 
process~\cite{PRC90.024001}, we expect that predictions calculated with 
other realistic NN potentials would not differ significantly.

The calculations have been performed including
all partial wave states with the total angular
momentum $ j \le 4$. 
In order to achieve fully converged results, 
60 $E_\gamma$ points and 50 $\theta_p$ points are used.

\begin{figure}
\includegraphics[width=7cm]{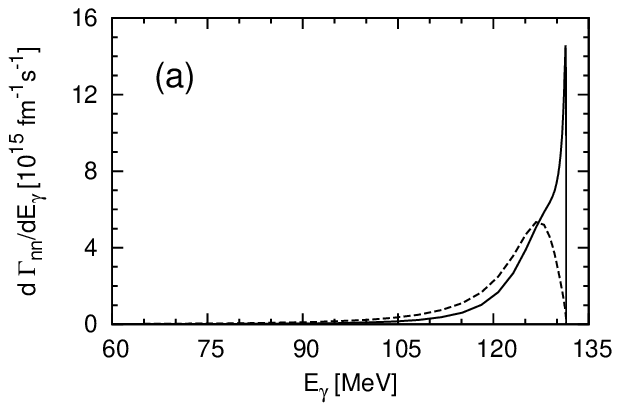}
\includegraphics[width=7cm]{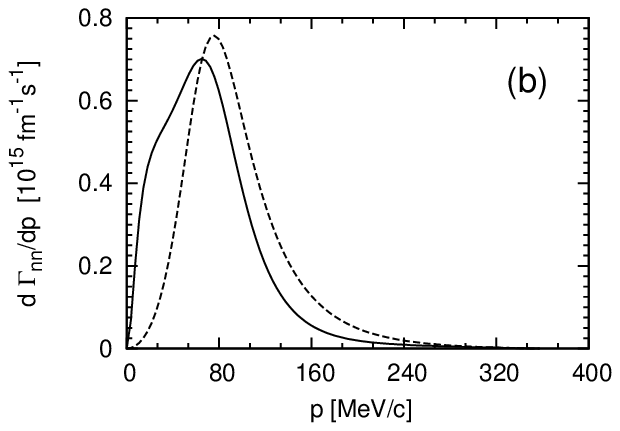}
\caption{The differential capture rate $ {d\Gamma_{nn}}/ {dE_\gamma} $
as a function of the photon energy $E_\gamma$
(a) and the differential capture rate
$ {d\Gamma_{nn}}/ {dp} $
as a function of the magnitude of the relative nn momentum $p$
(b)
for the $\pi^- + {^2{\rm H}}\rightarrow \gamma +n +n$ process,
calculated with the AV18 potential~\cite{av18}
and using the transition operator from Eq.~(\ref{jKR2}).
The dashed curves show the plane wave results and the solid curves
are used for the full results.
\label{Gamma_nn}}
\end{figure}

From Eq.~(\ref{gnn2}), one can easily extract the differential capture rate 
$ d\Gamma_{nn} /d E_\gamma $. This quantity is shown in the left panel of Fig.~\ref{Gamma_nn}
for the plane wave part of $N_{\pm 1}(m_1, m_2, m_d \, ) $ ((PW) dashed line)
and for the full $N_{\pm 1}(m_1, m_2, m_d \, ) $ ((Full) solid line).
When the full result for $N_{\pm 1}(m_1, m_2, m_d \, ) $ is taken,
a very narrow peak in the vicinity of $E_\gamma^{max,nn}$ emerges. 
Here, the relative energy of the two-neutron system
is very small, which explains strong rescattering effects in 
$ d\Gamma_{nn} /d E_\gamma $. 
Another form of the differential capture rate, $ d\Gamma_{nn} /d p $,
 is displayed
in the right panel of Fig.~\ref{Gamma_nn}, now 
as a function of the magnitude of the relative nn momentum.
The transition between $ d\Gamma_{nn} /d E_\gamma $
and $ d\Gamma_{nn} /d p $ reads 
\begin{eqnarray}
\frac{ d\Gamma_{nn} }{ d p } = 
\frac { 4 p } { E_\gamma + 2 M_n } \, \frac{ d\Gamma_{nn} }{d E_\gamma } \, .
\label{dgdf2}
\end{eqnarray}  

Despite the fact that the shapes of the differential rates, 
shown in Fig.~\ref{Gamma_nn} for the plane wave and full dynamics
are quite different, the corresponding integrated results 
for the total capture rate are rather similar.
We obtain 
$\Gamma_{nn}$= 0.318 $\times 10^{15} $~1/s (PW) 
and 
$\Gamma_{nn}$= 0.328 $\times 10^{15} $~1/s (Full).
Results of earlier calculations are displayed in Table~\ref{tab1}
and discussed in Sec.~\ref{section7}.

The total capture rate $\Gamma_{nn}$ can be also evaluated 
using other variables
\begin{eqnarray}
\Gamma_{nn} = \frac12 \frac1{( 2 \pi )^2 } \, 
\frac{ 2 \pi \alpha }{f_\pi^2 M_\pi } \,
\frac { \left(  M^\prime_d \alpha \, \right)^3 } {\pi  } \, 
\int\limits_{0}^{\pi} d \theta_{p_\gamma}  \sin \theta_{p_\gamma}  \, \int\limits_{0}^{2 \pi} d \phi_{p_\gamma}  \, 
\nonumber \\
\int\limits_{0}^{\pi} d \theta_{p_1}  \sin \theta_{p_1}  \, \int\limits_{0}^{2 \pi} d \phi_{p_1}  \, 
\int\limits_0^{E_1^{max}} \, dE_1 \,
\frac{M_n^2 p_1  E_\gamma    }{ E_\gamma + M_n + p_1 \cos \theta_{\gamma 1}    } \,
\nonumber \\
\frac13 \sum\limits_{m_d} \sum\limits_{m_1, m_2} 
\Big( 
\left| N_{+1}(m_1, m_2, m_d \, ) \, \right|^2  
\, + \, 
\left| N_{-1}(m_1, m_2, m_d \, ) \, \right|^2  \, 
\Big) \, ,
\label{gnn3}
\end{eqnarray}  
where 
$E_\gamma$ is the only physical solution of the nonrelativistic equation 
\begin{eqnarray}
E_\gamma^2 + 2 \left( M_n + p_1 \cos\theta_{\gamma 1} \right)  E_\gamma 
+ 2 \left( p_1^2 - M_n ( M_d + M_\pi - 2 M_n ) \, \right) = 0 \, ,
\label{Egamma1}
\end{eqnarray}
and depends on the magnitude of the detected neutron momentum, $p_1$, 
as well as on the angle between the detected neutron 
and photon momentum, $\theta_{\gamma 1}$.
Note that the maximal neutron energy $E_1^{max}$, which equals $\frac12 \left( M_d + M_\pi - 2 M_n \,  \right) $, 
does not depend 
on $\theta_{\gamma 1}$. Like Eq.~(\ref{gnn1}), also Eq.~(\ref{gnn3}) can be simplified, choosing 
$ {\hat {\bf p}}_\gamma = - {\hat {\bf z}} $ and the azimuthal angle of the neutron momentum $\phi_1=0$.

The building block of Eq.~(\ref{gnn3}) is the differential capture rate 
\begin{eqnarray}
d^5\Gamma_{nn}/\left( d\hat{\bf p}_\gamma d\hat{\bf p}_1 {dE_1} \right)
= \frac12 \frac1{( 2 \pi )^2 } \, 
\frac{ 2 \pi \alpha }{f_\pi^2 M_\pi } \,
\frac { \left(  M^\prime_d \alpha \, \right)^3 } {\pi  } \, 
\frac{M_n^2 p_1  E_\gamma    }{ E_\gamma + M_n + p_1 \cos \theta_{\gamma 1}    } \,
\nonumber \\
\frac13 \sum\limits_{m_d} \sum\limits_{m_1, m_2} 
\Big( 
\left| N_{+1}(m_1, m_2, m_d \, ) \, \right|^2  
\, + \, 
\left| N_{-1}(m_1, m_2, m_d \, ) \, \right|^2  \, 
\Big) \, .
\label{gnnde1}
\end{eqnarray}  
As early as in 1951 Watson and Stuart \cite{Watson1951} showed 
with quite simple dynamics that the corresponding photon spectrum is very 
sensitive to the properties of the low-energy nn interaction. 
Since then many calculations 
\cite{McVoy1961,Bander1964,Gibbs1975,Gibbs1977,deTeramond1977,deTeramond1980,deTeramond1987,
Gardestig2006PRC1,Gardestig2006PRL,Chen2008},
summarized in Ref.~\cite{Gardestig2009}, have also demonstrated 
that this capture process can be used 
to study the properties of low-energy 
nn scattering.

\begin{figure}
\includegraphics[width=9cm]{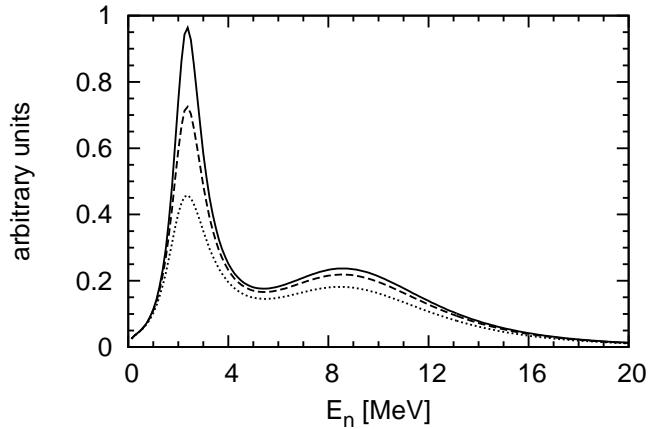}
\caption{The differential capture rate $ {d^5\Gamma_{nn}}/( d\hat{\bf p}_\gamma d\hat{\bf p}_1 {dE_1}) $
calculated with the AV18 potential~\cite{av18}
and using the transition operator from Eq.~(\ref{jKR2}) 
as a function of the neutron energy $E_n \equiv E_1$ for three different 
angles between the emitted photon and neutron momentum $\theta_{\gamma 1}$:
179$^\circ$ (solid line), 
175$^\circ$ (dashed line)
and 171$^\circ$ (dotted line). All the results correspond
to $a_{nn}$= -18.8~fm.
\label{Gamma_nn2}}
\end{figure}

\begin{figure}
\includegraphics[width=9cm]{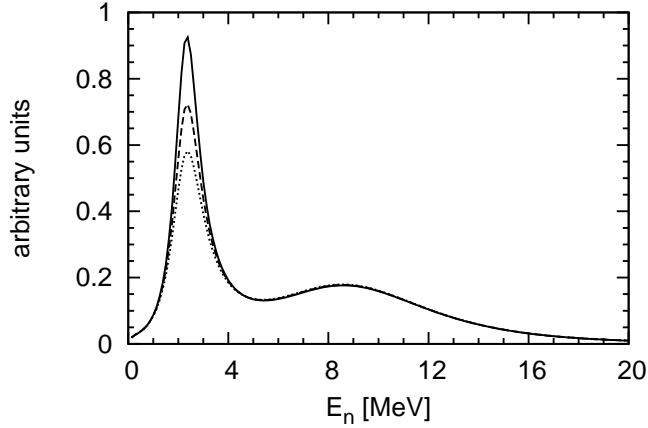}
\caption{The differential capture rate $ {d^5\Gamma_{nn}}/( d\hat{\bf p}_\gamma d\hat{\bf p}_1 {dE_1}) $
as a function of the neutron energy $E_n \equiv E_1$
calculated with
the transition operator from Eq.~(\ref{jKR2}) 
and with three versions of the AV18 potential~\cite{av18}
yielding different $a_{nn}$ values: -21.8~fm (solid line),
-18.8~fm (dashed line) and -16.5~fm (dotted line).
All the results correspond to $\theta_{\gamma 1}$= 179$^\circ$.
\label{Gamma_nn3}}
\end{figure}

Also results of our calculations (see Figs.~\ref{Gamma_nn2}-\ref{Gamma_nn3})
show that the photon spectrum has two salient peaks: the nn
FSI peak around $E_n \equiv E_1= 2$~MeV and the so-called QFS peak arising from 
the quasi-free $\pi^-\, p$ process around $E_n= 9$~MeV.
From Fig.~\ref{Gamma_nn2}, it is clear that heights of the peaks increase 
with increasing $\theta_{\gamma 1}$.
What is more important, for fixed $\theta_{\gamma 1}$ small variations of the $^1S_0$ nn
interaction lead to quite visible changes in the FSI peak.
We study this effect, performing additional calculations 
with the altered nn AV18 potential whose
$^1S_0$ matrix elements are multiplied by the factor $1.01$ and $0.99$.
This leads to the following changes of the $a_{nn}$ values: 
for the stronger version of the potential we get $a_{nn}= -21.8$~fm, while 
for the weakened force $a_{nn}= -16.5$~fm. (The original value is $a_{nn}= -18.8$~fm.) 
The primary and modified neutron spectra are displayed in Fig.~\ref{Gamma_nn3}
and we see that the FSI peak becomes higher if the absolute value of $a_{nn}$ grows.
Note that in all these calculations the nn AV18 potential was used without electromagnetic
contributions.

\begin{figure}
\includegraphics[width=9cm]{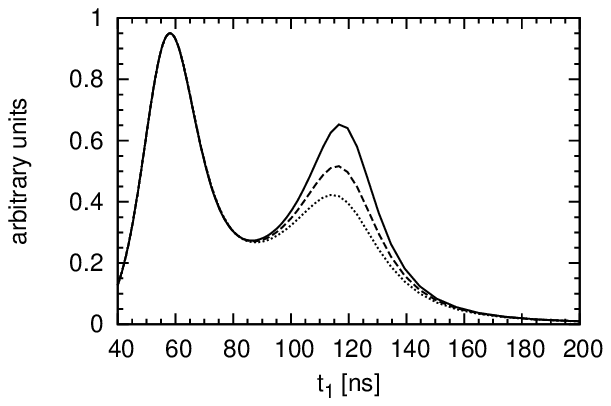}
\caption{The differential capture rate $ {d^5\Gamma_{nn}}/( d\hat{\bf p}_\gamma d\hat{\bf p}_1 {dt_1}) $
(neutron time-of-flight spectrum for a flight path of 2.55~m) 
obtained from the results shown in Fig.~\ref{Gamma_nn3}.
The results corresponding to different 
$a_{nn}$ values, represented by the same lines as in Fig.~\ref{Gamma_nn3}, are
 normalized at the (left) QFS peak. 
\label{Gamma_nn4}}
\end{figure}

Actually, in order to minimize systematic uncertainties in the $a_{nn}$ extraction, 
the very shape of the neutron time-of-flight spectrum in the area 
corresponding to the FSI peak is considered \cite{Schori1987,Chen2008}.
The nonrelativistic relation between the time-of-flight variable $t_1$ and 
the energy of the neutron $E_n$ reads 
\begin{eqnarray} 
E_n = \frac12 M_n s^2 \frac1{t_1^2} \, , 
\end{eqnarray}  
where $s$ is the flight path to the neutron detector.
A simple step then leads  to the neutron time-of-flight spectrum 
demonstrated in Fig.~\ref{Gamma_nn4} (for $s$= 2.55~m) for the same 
three values of $a_{nn}$:
\begin{eqnarray}
d^5\Gamma_{nn}/\left( d\hat{\bf p}_\gamma d\hat{\bf p}_1 {dt_1} \right)
= \frac12 \frac1{( 2 \pi )^2 } \, 
\frac{ 2 \pi \alpha }{f_\pi^2 M_\pi } \,
\frac { \left(  M^\prime_d \alpha \, \right)^3 } {\pi  } \, 
\frac{M_n p_1^3  E_\gamma    }{ t_1 \,  \left( E_\gamma + M_n + p_1 \cos \theta_{\gamma 1}  \right)  } \,
\nonumber \\
\frac13 \sum\limits_{m_d} \sum\limits_{m_1, m_2} 
\Big( 
\left| N_{+1}(m_1, m_2, m_d \, ) \, \right|^2  
\, + \, 
\left| N_{-1}(m_1, m_2, m_d \, ) \, \right|^2  \, 
\Big) \, .
\label{gnndt1}
\end{eqnarray} 
Equation (\ref{gnndt1}) can be compared with the relativistic formula 
from Eq.~(3) in Ref.~\cite{Gardestig2006PRC1}. 

We do not intend here to work on the extraction of $a_{nn}$ with our present theory. 
Before moving to calculations with 3N systems we wanted to make sure that 
our framework possesses the same important features as the calculations used 
in the old and more recent analyses of the $ \pi^- + {^2{\rm H}}  \rightarrow \gamma + n + n $ process 
\cite{McVoy1961,Bander1964,Gibbs1975,Gibbs1977,deTeramond1977,deTeramond1987,Gardestig2006PRC1,Chen2008}.

\section{Results for the $\pi^- + {^3{\rm He}} \rightarrow \gamma + {^3{\rm H}} $ reaction}
\label{section4}

In this case, we deal with two-body kinematics
and we can compare the photon energy calculated nonrelativistically 
and using relativistic equations. 
The relativistic result, based on 
\begin{eqnarray}
M_\pi + M_{{^3{\rm He}}} = E_\gamma + \sqrt{ E_\gamma^2 + M_{{^3{\rm H}}}^2 \, }
\label{3h1rel}
\end{eqnarray}
reads
\begin{eqnarray}
\left( E_\gamma \right)^{rel} = \frac{ \left( M_{{^3{\rm He}}} + M_\pi \, \right)^2 - M_{{^3{\rm H}}}^2   } 
{ 2 \, \left( M_{{^3{\rm He}}} + M_\pi \, \right)   } \, .
\label{3h2rel}
\end{eqnarray}
In the nonrelativistic case, we start with 
\begin{eqnarray}
M_\pi + M_{{^3{\rm He}}} = E_\gamma + M_{{^3{\rm H}}} + \frac{ E_\gamma^2 } { 2 M_{{^3{\rm H}}}  }
\label{3h1nrl}
\end{eqnarray}
and arrive at
\begin{eqnarray}
\left( E_\gamma \right)^{nrl} = - M_{{^3{\rm H}}} 
+ \sqrt{ M_{{^3{\rm H}}} \left( -M_{{^3{\rm H}}} + 2 \left( M_{{^3{\rm He}}} + M_\pi \, \right)  \right) }
 \, .
\label{3h2nrl}
\end{eqnarray}
Again the obtained numerical values,
$ \left( E_\gamma \right)^{rel}$  = 135.760 MeV
and
$ \left( E_\gamma \right)^{nrl}$  = 135.743 MeV,
are very close to each other.

For this reaction we calculate only the total capture rate
\begin{eqnarray}
&&\Gamma_{{^3{\rm H}}} = \frac12 \frac1{( 2 \pi )^2 } \, {\cal R} \, \frac{ 2 \pi \alpha }{f_\pi^2 M_\pi E_\gamma } \,
\frac { \left(  2 M^\prime_{{^3{\rm He}}} \alpha \, \right)^3 } {\pi  } \, \rho \, 
\nonumber \\
&&4 \pi \,
\frac12 \sum\limits_{m_{{^3{\rm He}}}} \sum\limits_{m_{{^3{\rm H}}} } 
\Big( 
\left| N_{+1} (m_{{^3{\rm H}}}, m_{{^3{\rm He}}}  \, ) \, \right |^2 \, + \, 
\left| N_{-1} (m_{{^3{\rm H}}}, m_{{^3{\rm He}}}  \, ) \, \right |^2 \, \Big)  \, ,
\label{g3h}
\end{eqnarray}  
where 
the factor $ \frac { \left( 2  M^\prime_{{^3{\rm He}}} \alpha \, \right)^3 } {\pi  } $, like 
in the deuteron case, comes from the $K$-shell atomic wave function 
and $ M^\prime_{{^3{\rm He}}}  = \frac { M_{{^3{\rm He}}} M_\pi } { M_{{^3{\rm He}}} + M_\pi }$. 
For the unpolarized reaction the angular integration over the photon momenta 
leads to the $ 4 \pi$ factor and in all the considered cases we set $ {\hat {\bf p}}_\gamma = - {\hat {\bf z}} $.
The phase space factor $ \rho$ is 
\begin{eqnarray}
\rho = \frac {E_\gamma^2 } {1 + \frac {E_\gamma}{\sqrt{ E_\gamma^2 + M_{{^3{\rm H}}}^2 \, } }  } \, 
\approx \,
E_\gamma^2 \,  \left( 1 - \frac {E_\gamma} {M_{{^3{\rm H}}} } \,   \right)  \, .
\label{rho3h}
\end{eqnarray}  
Like for muon capture, the additional factor ${\cal R}$ accounts for the finite 
volume of the $^3$He charge and we use the same value ${\cal R}= 0.98 $ as in Ref.~\cite{prc83.014002}.  
(The corresponding correction in the deuteron case has been neglected.)
The nuclear matrix elements involve the initial $^3$He and final $^3$H states:
\begin{eqnarray}  
N_{\pm 1} (m_{{^3{\rm H}}}, m_{{^3{\rm He}}}  \, )  \, \equiv \, 
\BA{\Psi_{{^3{\rm H}}} \, {\bf P}_{\!f}=-{\bf p}_\gamma \,\, m_{{^3{\rm H}}} \, } \, 
j_{A, \pm 1}
\, \KT{\Psi_{{^3{\rm He}}} \, {\bf P}_{\!i}=0 \,\, m_{{^3{\rm He}}} \, } 
\label{n3h}
\end{eqnarray}  
and are obtained employing our standard PWD techniques 
\cite{physrep, PRC90.024001}.

Our results for this process are  obtained for two cases.
When we generate the $^3$He and $^3$H wave functions 
using the AV18 NN potential only, we get
$\Gamma_{^3{\rm H}}$= 2.059 $\times 10^{15} $~1/s.
For the wave functions calculated with the AV18 NN potential augmented 
by the Urbana~IX 3N force, 
the rate is slightly reduced to $\Gamma_{^3{\rm H}}$= 2.013 $\times 10^{15} $~1/s.
These predictions are compared with the results of earlier 
calculations in Table~\ref{tab2} and discussed in Sec.~\ref{section7}.

\section{Results for the $\pi^- + {^3{\rm He}} \rightarrow \gamma + n + d $ 
and $\pi^- + {^3{\rm He}} \rightarrow \gamma + n + n + p $ reactions}
\label{section5}

\begin{figure}
\includegraphics[width=7cm]{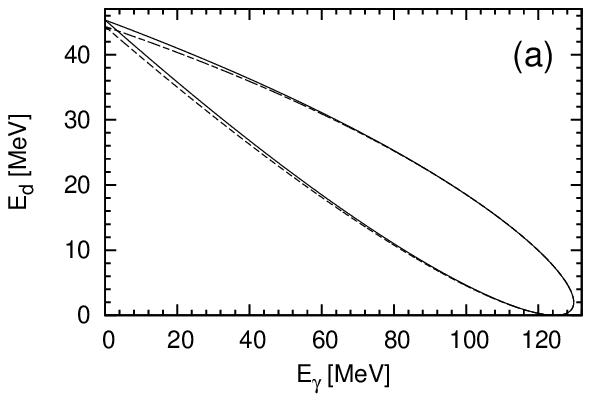}
\includegraphics[width=7cm]{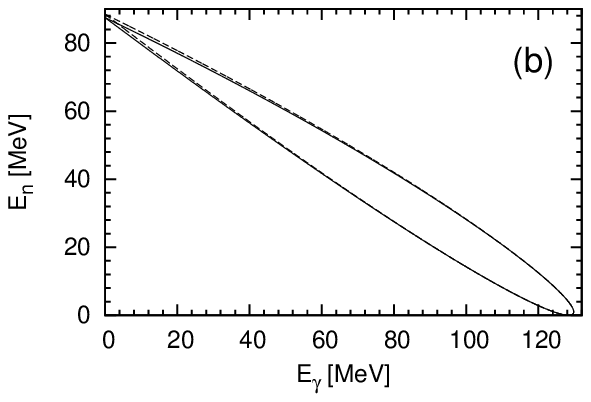}
\caption{The kinematically allowed region in the
$E_\gamma - E_d$ (a) 
and 
$E_\gamma - E_n$ (b) plane
calculated relativistically
(solid curve) and
nonrelativistically (dashed curve) for the
$\pi^-+ {^3{\rm He}} \rightarrow \gamma + n + d $ process. 
The lines overlap except for small photon energies.
\label{egamma_3He_ed}}
\end{figure}

\begin{figure}
\includegraphics[width=8cm]{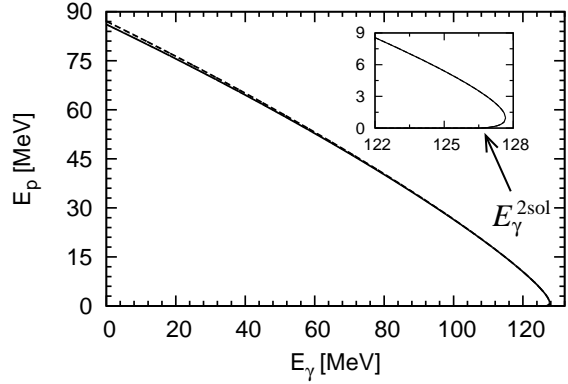}
\caption{The kinematically allowed region in the
$E_\gamma - E_p$ plane calculated relativistically
(solid curve) and
nonrelativistically (dashed curve) for the
$\pi^-+ {^3{\rm He}} \rightarrow \gamma + n + n + p $ process.
The inset focuses on the highest photon energy region.
The lines practically overlap except for very small photon energies.
\label{egamma_3He_ep}}
\end{figure}

The kinematics of the 
$\pi^-+ {^3{\rm He}} \rightarrow \gamma + n + d $ 
and
$\pi^-+ {^3{\rm He}} \rightarrow \gamma + n + n + p $ 
reactions is treated exactly in the same way
as in muon capture on $^3$He \cite{PRC90.024001},
so we can immediately evaluate
the maximal photon energies for the two breakup channels as
\begin{eqnarray}
\left( E_\gamma^{max,nd} \, \right)^{rel} &=&
\frac{({M_{{^3{\rm He}}}}-{M_d}+{M_\pi}-{M_n})
   ({M_{{^3{\rm He}}}}+{M_d}+{M_\pi}+
   {M_n})}{2 ({M_{{^3{\rm He}}}}+{M_\pi})} \, ,
\label{Enumaxrelnd} \\
\left( E_\gamma^{max,nnp} \, \right)^{rel} &=&
\frac{{M_{{^3{\rm He}}}}^2+2 {M_{{^3{\rm He}}}}
   {M_\pi}+{M_\pi}^2-(2
   {M_n}+{M_p})^2}{2
   ({M_{{^3{\rm He}}}}+{M_\pi})} \, ,
\label{Enumaxrelnnp} \\
\left( E_\gamma^{max,nd} \, \right)^{nrl} &=&
\sqrt{
( M_d + M_n ) ( 2 M_{{^3{\rm He}}} + 2 M_\pi - M_d - M_n ) 
}-{M_d}-{M_n} \, ,
\label{Enumaxnrlnd} \\
\left( E_\gamma^{max,nnp} \, \right)^{nrl} &=&
\sqrt{
( M_p + 2 M_n ) ( 2 M_{{^3{\rm He}}} + 2 M_\pi - 2 M_n - M_p ) 
}-2 {M_n}-{M_p} \, .
\label{Enumaxnrlnnp}
\end{eqnarray}
The numerical values are the following:
$\left( E_\gamma^{max,nd} \, \right)^{rel} $ = 129.794 MeV,
$\left( E_\gamma^{max,nd} \, \right)^{nrl} $ = 129.792 MeV,
$\left( E_\gamma^{max,nnp} \, \right)^{rel} $ = 127.668 MeV
and
$\left( E_\gamma^{max,nnp} \, \right)^{nrl} $ = 127.667 MeV.
The kinematically allowed regions 
in the $E_\gamma - E_d$ 
and
in the $E_\gamma - E_n$ 
planes for the 
two-body breakup of $^3$He are shown in Fig.~\ref{egamma_3He_ed}.
For both cases, lines obtained with the relativistic and nonrelativistic kinematics
fully overlap except for very small photon energies.
The same is also true for the three-body breakup, as demonstrated
in Fig.~\ref{egamma_3He_ep} for the allowed region
in the $E_\gamma - E_p$ plane. 
In this case the minimal proton kinetic energy is greater than zero
for $E_\gamma > E_\gamma^{2sol}$ (see the inset in Fig.~\ref{egamma_3He_ep})
and the  values of $E_\gamma^{2sol}$
based on the relativistic kinematics, 
\begin{equation}
\left( E_\gamma^{2sol} \, \right)^{rel} =
\frac{
( M_{{^3{\rm He}}} + M_\pi ) ( M_{{^3{\rm He}}} + M_\pi - 2 M_p ) - 4 {M_n}^2+{M_p}^2
}{2 ({M_{{^3{\rm He}}}}+{M_\pi}-{M_p})}
\label{Enu2solrel}
\end{equation}
and nonrelativistic one,
\begin{equation}
\left( E_\gamma^{2sol} \, \right)^{nrl} =
2 \left(\sqrt{{M_{{^3{\rm He}}}}
   {M_n}+{M_\pi}
   {M_n}-{M_n}^2-
   {M_n}
   {M_p}}-{M_n}\right) \, ,
\label{Enu2solnrl}
\end{equation}
yield very similar numerical values:
126.318 MeV
and
126.314 MeV, respectively. All these results clearly show that 
the nonrelativistic kinematics can be safely used also for the breakup 
channels.

Using standard steps we obtain the formulas for the total capture rates.
In the case of the two-body breakup it reads:
\begin{eqnarray}
&&\Gamma_{nd} = \frac12 \frac1{( 2 \pi )^2 } \, \frac{ 2 \pi \alpha }{f_\pi^2 M_\pi } \,
{\cal R} \, \frac { \left(  2 M^\prime_{{^3{\rm He}}} \alpha \, \right)^3 } {\pi  } \, 4 \pi \, 
\nonumber \\
&&\int\limits_0^{E_\gamma^{max,nd}} \, dE_\gamma E_\gamma  \,
\frac23 M q_0  \, \frac13 \, 
\int\limits_{0}^{\pi} d \theta_{q_0} \sin \theta_{q_0} \, 2 \pi \, 
\nonumber \\
&&\frac12 \sum\limits_{m_{{^3{\rm He}}}} \sum\limits_{m_{n} , m_d } 
\Big( 
\left| N_{nd, \, +1} (m_{n}, m_d, m_{{^3{\rm He}}}  \, ) \, \right |^2 \, + \, 
\left| N_{nd, \, -1} (m_{n}, m_d, m_{{^3{\rm He}}}  \, ) \, \right |^2 \, 
\Big) \, ,
\label{gnd}
\end{eqnarray}  
where we used the relative neutron-deuteron momentum
\begin{eqnarray}
{\bf q}_0  \equiv \frac23 \left(  {\bf p}_n - \frac12 {\bf p}_d \,  \right)  \, ,
\label{q0}
\end{eqnarray}
given in terms of the final neutron (${\bf p}_n$) and deuteron (${\bf p}_d$) momenta,
to evaluate $\Gamma_{nd}$.
For the $ \pi^- + {^3{\rm He}} \rightarrow \gamma + n + n + p $ reaction 
we obtain in a similar way:
\begin{eqnarray}
&&\Gamma_{nnp} = \frac12 \frac1{( 2 \pi )^2 } \, \frac{ 2 \pi \alpha }{f_\pi^2 M_\pi } \,
{\cal R} \, \frac { \left(  2 M^\prime_{{^3{\rm He}}} \alpha \, \right)^3 } {\pi  } \, 4 \pi \, 
\nonumber \\
&&\int\limits_0^{E_\gamma^{max,nnp}} \, dE_\gamma E_\gamma  \,
\frac13 \, 
\int\limits_{0}^{\pi} d \theta_{q} \sin \theta_{q} \, 2 \pi \, 
\int\limits_{0}^{\pi} d \theta_{p} \sin \theta_{p} \, \int\limits_{0}^{2 \pi} d \phi_{p} \, 
\int\limits_0^{p^{max}} \, dp p^2  \, \frac23 M q  \, 
\nonumber \\
&&\frac12 \sum\limits_{m_{{^3{\rm He}}}} \sum\limits_{m_1, m_2 , m_p } 
\Big( 
\left| N_{nnp, \, +1} (m_1, m_2, m_p, m_{{^3{\rm He}}}  \, ) \, \right |^2 \, + \, 
\left| N_{nnp, \, -1} (m_1, m_2, m_p, m_{{^3{\rm He}}}  \, ) \, \right |^2 \, 
\Big) \, .
\label{gnnp}
\end{eqnarray}  
Here the integral is expressed in terms of the Jacobi relative 
momenta ${\bf p}$ and ${\bf q}$, that is  
\begin{eqnarray}
{\bf p}  \equiv \frac12 \left(  {\bf p}_1 - {\bf p}_2 \,  \right) \, , \nonumber \\
{\bf q}  \equiv \frac23 \left(  {\bf p}_p - \frac12 \left(  {\bf p}_1 +  {\bf p}_2 \,  \right) \,  \right) \, ,
\label{pq}
\end{eqnarray} 
obtained from the proton momentum (${\bf p}_p$) and the momenta of the two neutrons (${\bf p}_1$ and  ${\bf p}_2$).
In Eq.~(\ref{gnnp}) $p^{max}$ is a function of $E_\gamma$ 
and $q \equiv \left| {\bf q} \right| = q (E_\gamma , p \, )$~\cite{PRC94.034002}.
Note that we used the same geometrical arguments as before
to simplify the angular integrations in Eqs.~(\ref{gnd}) and (\ref{gnnp}).

The crucial matrix elements
\begin{eqnarray}  
N_{nd, \pm 1 } (m_n, m_d , m_{{^3{\rm He}}}  \, )  \, \equiv \, 
\BA{\Psi_{nd}^{(-)}  \, {\bf P}_{\!f}=-{\bf p}_\gamma \, m_n \, m_d } \, 
j_{A, \pm 1}
\, \KT{\Psi_{{^3{\rm He}}} \, {\bf P}_{\!i}=0 \, m_{{^3{\rm He}}} \, } 
\label{nnd}
\end{eqnarray}  
and
\begin{eqnarray}  
N_{nnp, \pm 1 } (m_1, m_2 , m_p , m_{{^3{\rm He}}}  \, )  \, \equiv \, \nonumber \\
\BA{\Psi_{nnp}^{(-)}  \, {\bf P}_{\!f}=-{\bf p}_\gamma \, m_1 \, m_2 \, m_p} \, 
j_{A, \pm 1}
\, \KT{\Psi_{{^3{\rm He}}} \, {\bf P}_{\!i}=0 \, m_{{^3{\rm He}}} \, } 
\label{nnnp}
\end{eqnarray}  
are calculated in momentum space, as outlined in Ref.~\cite{PRC90.024001},
within the numerical framework developed in Refs.~\cite{physrep,romek2}. 
Also in Ref.~\cite{physrep} the detailed definitions of various 3N dynamics
can be found.

\begin{figure}
\includegraphics[width=8cm]{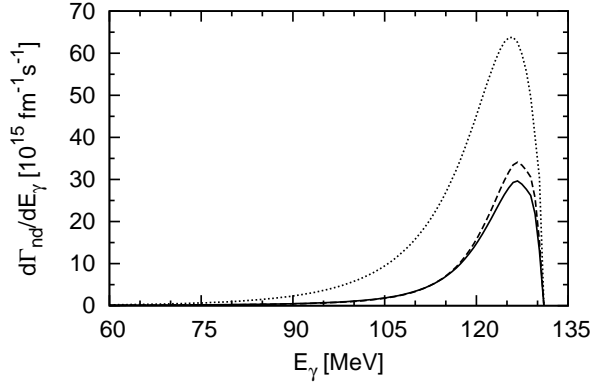}
\caption{The differential capture rate $ {d\Gamma }_{nd}/ {dE_\gamma} $
for the $ \pi^- + ^{3}{\rm He} \rightarrow \gamma + n + d$ process
as a function of the photon energy,
calculated with the
single nucleon transition operator and with different treatment of 3N dynamics:
taking the symmetrized plane wave approximation (dotted line),
calculating the initial and final 3N states without (dashed curve)
and with 3N force (solid line).
The calculations are based on the AV18 NN potential \cite{av18}
and the Urbana~IX 3N force \cite{urbana}.
\label{nddynamics}}
\end{figure}

Since the convergence of our PWD-based results 
with respect to the total subsystem angular momentum $j$ 
and the total 3N angular momentum $J$ was discussed in Ref.~\cite{PRC90.024001},
we can start the discussion of our predictions with Fig.~\ref{nddynamics}, 
where, for the 
$ \pi^- + {^3{\rm He}} \rightarrow \gamma + n + d$ reaction,
we compare results of calculations employing various 3N dynamics:
symmetrized plane wave approximation obtained with the AV18 NN potential,
consistent calculations of the initial and final nuclear states 
with the AV18 interaction only
and calculations based on the Hamiltonian containing 
additionally the Urbana~IX 3N force. 
The results are qualitatively quite similar to the ones obtained for muon capture.
The three differential capture rates 
$d \Gamma_{nd}/dE_{\gamma}$ rise very slowly with the photon energy and
form a single maximum close to the maximal photon energy.
This maximum is higher and broader 
for the plane wave case. Effects introduced by FSI
are very important and, in the maximum, reduce the full $d \Gamma_{nd}/dE_{\gamma}$ to about $1/2$ 
of the plane wave prediction. 
The inclusion of the 3N force lowers 
the peak further by about $14\ $\% .

\begin{figure}
\includegraphics[width=8cm]{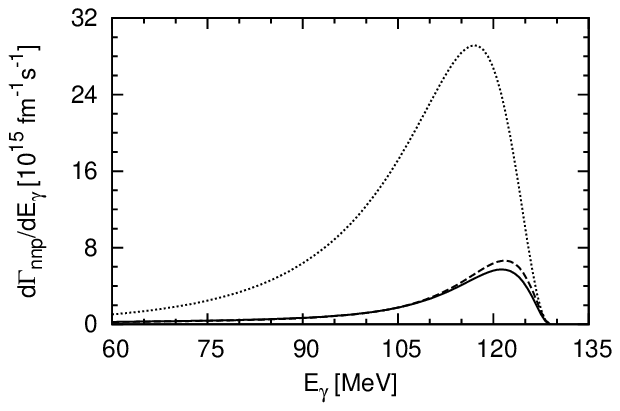}
\caption{The same as in Fig.~\ref{nddynamics}
for the differential capture rate $ {d\Gamma }_{nnp}/ {dE_\gamma} $
in the case of the $ \pi^- + {^3{\rm He}} \rightarrow \gamma + n + n + p$ process.
\label{nnpdynamics}}
\end{figure}

The FSI effects are even stronger 
for the $ \pi^- + {^3{\rm He}} \rightarrow \gamma + n + n + p$ reaction,
as displayed in Fig.~\ref{nnpdynamics} for the differential rate
$d \Gamma_{nnp}/dE_{\gamma}$.
The reduction factor is already about 4.5 for the case without 3N force.
The 3N force has roughly the same effect as in the two-body breakup case.
In Ref.~\cite{PRC90.024001} we suggested that this might be a consequence
of the overprediction of the $A=3$ radii when 3N interaction is neglected.

\begin{figure}
\includegraphics[width=8cm]{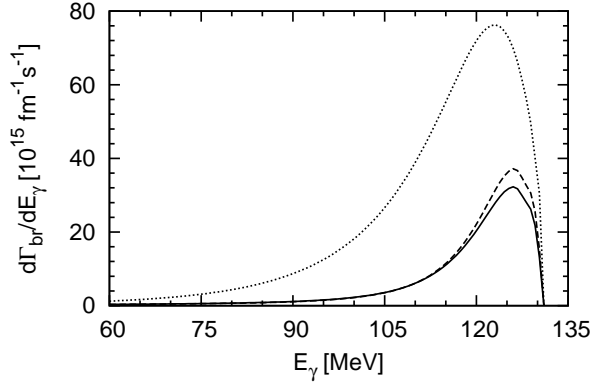}
\caption{The same as in Fig.~\ref{nddynamics}
for the differential breakup capture rate
$ {d\Gamma }_{br}/ {dE_\gamma} =
  {d\Gamma }_{nd}/ {dE_\gamma} +
  {d\Gamma }_{nnp}/ {dE_\gamma} $.
\label{brdynamics}}
\end{figure}

\begin{figure}
\includegraphics[width=8cm]{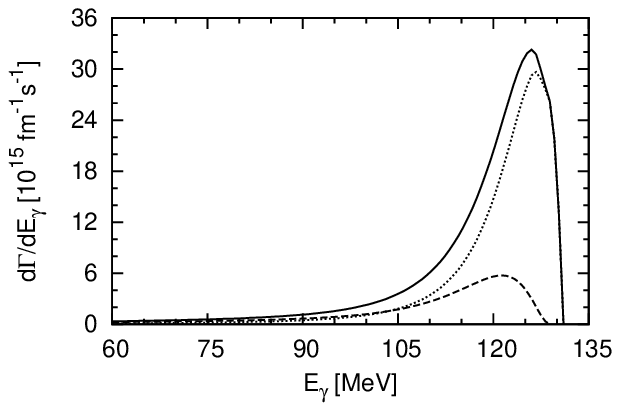}
\caption{Two-body (dotted curve) and three-body breakup contributions (dashed curve)
to the differential breakup radiative pion capture rate (solid curve)
as a function of the photon energy,
calculated with the
single nucleon transition operator and with full 3N dynamics
(including 3N force in the initial and final nuclear states).
As before, the calculations are based on the AV18 NN potential \cite{av18}
and the Urbana~IX 3N force \cite{urbana}.
\label{ndandnnpcontributions}}
\end{figure}

Since the values of $d \Gamma_{nnp}/dE_{\gamma}$ are much smaller (by a factor of 5) 
than $d \Gamma_{nd}/dE_{\gamma}$, at least in the peak area, 
the picture shown in Fig.~\ref{brdynamics} for the total breakup capture rate,
$d \Gamma_{br}/dE_{\gamma} = d \Gamma_{nd}/dE_{\gamma} + d \Gamma_{nnp}/dE_{\gamma}$,
is similar to the one for $d \Gamma_{nd}/dE_{\gamma}$.  
Finally, in Fig.~\ref{ndandnnpcontributions}, we show the contributions
from the two-body and three-body breakup channels calculated with our full dynamics,
that is including the 3N force. This figure clearly demonstrates 
that the breakup is dominated by the two-body channel.
The corresponding predictions for the total $\Gamma_{nd}$ 
and 
$\Gamma_{nnp} $ capture rates 
are presented in Table~\ref{tab2} 
together with earlier theoretical predictions 
and experimental information about the 
relative probability of the breakup and non-breakup radiative capture.
These results will be discussed in Sec.~\ref{section7}.

\section{Results for the $\pi^- + {^3{\rm H}} \rightarrow \gamma + n + n + n$ reaction}
\label{section6}

\begin{figure}
\begin{center}
\includegraphics[width=9cm]{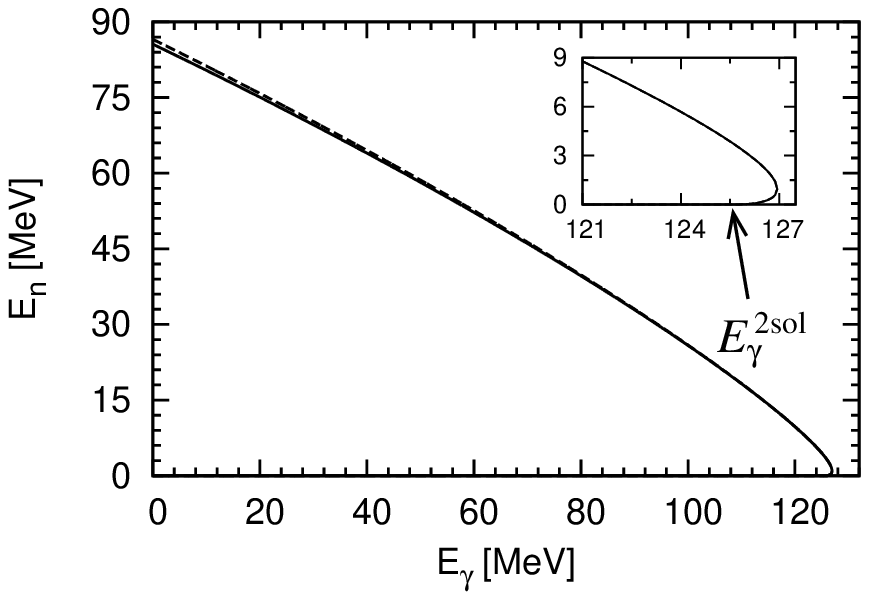}
\end{center}
\caption{The kinematically allowed region in the
$E_\gamma - E_n$ plane calculated relativistically
(solid curve) and
nonrelativistically (dashed curve) for the
$\pi^- + {^3{\rm H}} \rightarrow \gamma + n + n + n $ capture process. \label{egamma_3H_en}}
\end{figure}

The kinematically allowed region
in the $E_\gamma - E_n$ plane for the
three-body breakup of $^3$H is shown in Fig.~\ref{egamma_3H_en}.
As for the $\pi^- + {^3{\rm He}} \rightarrow \gamma + n + n + p$ capture process,
we show the border lines based on the relativistic and nonrelativistic kinematics
and evaluate correspondingly the maximal photon energy
relativistically 
$\left( E_\gamma^{max,nnn} \, \right)^{rel} $ = 126.940 MeV
and nonrelativistically 
$\left( E_\gamma^{max,nnn} \, \right)^{nrl} $ = 126.939 MeV.
For completeness we give also values of $E_\gamma^{2sol} $:
$\left( E_\gamma^{2sol} \, \right)^{rel} $= 125.604 MeV
and
$\left( E_\gamma^{2sol} \, \right)^{nrl} $= 125.600 MeV.
As expected, the kinematics of this reaction can be described using 
the nonrelativistic formulas in the nuclear sector, consistent with
the nonrelativistic dynamics.

In the calculations of the three-neutron continuum only
the nn version of the AV18 potential \cite{av18}
appears. The nuclear Hamiltonian contains the same Urbana~IX 
3N force \cite{urbana}. 
The formula for the total $\Gamma_{nnn}$ capture rate,
\begin{eqnarray}
&&\Gamma_{nnn} = \frac12 \frac1{( 2 \pi )^2 } \, \frac{ 2 \pi \alpha }{f_\pi^2 M_\pi } \,
\frac { \left( M^\prime_{{^3{\rm H}}} \alpha \, \right)^3 } {\pi  } \, 4 \pi \, 
\nonumber \\
&&\int\limits_0^{E_\gamma^{max,nnn}} \, dE_\gamma E_\gamma  \,
\frac23 M q  \, \frac19 \, 
\int\limits_{0}^{\pi} d \theta_{q} \sin \theta_{q} \, 2 \pi \, 
\int\limits_{0}^{\pi} d \theta_{p} \sin \theta_{p} \, \int\limits_{0}^{2 \pi} d \phi_{p} \, 
\int\limits_0^{p^{max}} \, dp p^2  \,
\nonumber \\
&&\frac12 \sum\limits_{m_{{^3{\rm H}}}} \sum\limits_{m_1, m_2 , m_3 } 
\Big( 
\left| N_{nnn, \, +1} (m_1, m_2, m_3, m_{{^3{\rm H}}}  \, ) \, \right |^2 \, + \, 
\left| N_{nnn, \, -1} (m_1, m_2, m_3, m_{{^3{\rm H}}}  \, ) \, \right |^2 \, 
\Big) \, ,
\label{gnnn}
\end{eqnarray}
is a modification of Eq.~(\ref{gnnp}), taking into 
account that, like for the deuteron case, $Z=1$, ${\cal R}=1$, 
and that there are three identical particles in the final state.

\begin{figure}
\includegraphics[width=8cm]{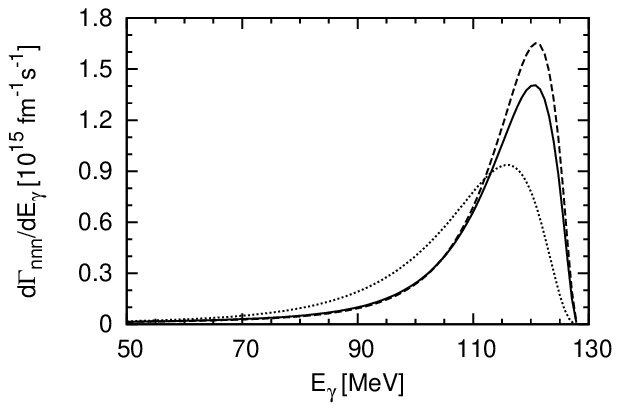}
\caption{The same as in Fig.~\ref{nddynamics}
for the differential capture rate $ {d\Gamma }_{nnn}/ {dE_\gamma} $
in the case of the $ \pi^- + {^3{\rm H}} \rightarrow \gamma + n + n + n$ process.
\label{brdynamics2}}
\end{figure}

The values of the differential capture rates 
$d \Gamma_{nnn}/dE_{\gamma}$ are smaller (see Fig.~\ref{brdynamics2}) than
the $d \Gamma_{nnp}/dE_{\gamma}$ results
for the three-body breakup of $^3$He.
The maximum is still broader for the plane wave case 
but FSI 
now {\em raises} the results by a factor of 1.8,
playing a crucial role also for this process.
The inclusion of the 3N force leads to a
reduction of the peak's height by about $15\ $\%.
Also the total rates for pion capture on $^3$H are 
displayed in Table~\ref{tab2} and described in the next section.

\section{Comparison with earlier theoretical predictions}
\label{section7}

A comparison with earlier theoretical predictions and experimental 
data is not simple for several reasons. Experimental work never 
actually aimed at obtaining total radiative capture rates.
Very often measurements covered different reaction channels and tried
to gather information about their relative probabilities. The famous Panofsky 
ratio for $^3$He, studied in Refs.~\cite{Phillips1974,Gibbs1978}, is a very good example. Unfortunately, such ratios 
involve rates, which are not calculated by us in the present paper.
Also we do not have  experimental (unnormalized) photon 
energy spectra for the reactions of interest at our disposal.

Many calculations for the $\pi^- + {^2{\rm H}} \rightarrow \gamma + n + n $
concentrated on the extraction of the nn scattering length 
and their authors did not provide results for the 
total radiative capture $\Gamma_{nn}$.
This is because already Gibbs, Gibson and Stephenson pointed out in Ref.~\cite{Gibbs1977}
that the total radiative rate is clearly insensitive to uncertainties of
the low-energy nn scattering parameters. Namely, they observed
that variation of $a_{nn}$ from $-15$ to $-20$~fm or the effective 
range $r_{nn}$ from $2.6$ to $3.0$~fm changed $\Gamma_{nn}$ by less than 1~\%.
Thus predictions for the $\Gamma_{nn}$ capture rate
can be found in only few theoretical papers~\cite{Reitan1966,Sotona1976,Gibbs1977}.

Additional problems are caused by the fact that some predictions 
originate from combined theoretical and experimental evidence.
Finally, earlier calculations are often corrected in subsequent 
publications. Nonetheless, we have tried to collect the available 
information and we show it in Tables~\ref{tab1} and \ref{tab2}.

In Table~\ref{tab1} 
the total radiative capture rates
for the $ \pi^- + {^2{\rm H}} \rightarrow \gamma + n + n $
reaction are displayed. As already mentioned, for this observable our plane wave 
and full results are rather similar despite the fact that final
state rescattering is very important and 
substantially affects the differential rates in Figs.~\ref{Gamma_nn} and \ref{Gamma_nn2}.
We notice also that our full predictions agree with the earlier
theoretical results, except for Ref.~\cite{Gibbs1977}.

Table~\ref{tab2} contains rates for radiative capture in the trinucleons.
We consider several capture channels, starting from the only non-breakup 
process: $\pi^- + {^3{\rm He}} \rightarrow \gamma + {^3{\rm H}}$.
Here the values of the rates are much higher than for the  
$ \pi^- + {^2{\rm H}} \rightarrow \gamma + n + n $ reaction. 
The capture rate is raised by approximately 3.5~\%, when
the 3N bound states are calculated consistently not only with 
the 2N but also with the 3N potentials.

The breakup of $^3$He is clearly dominated by the 
$\pi^- + {^3{\rm He}} \rightarrow \gamma + n + d$ reaction,
since the rate for the 
$\pi^- + {^3{\rm He}} \rightarrow \gamma + n + n + p$
process is three times smaller. For both breakup
reactions FSI effects based on the 2N forces 
are very strong and reduce the rates significantly. 
The inclusion of the 3N force leads to a further reduction,
which amounts to 9~\% (two-body breakup) and to 7~\% (three-body breakup).
Our best results (obtained with the 3N force) for the ratio
of the total breakup rate to the non-breakup rate 
agree both with the experimental data from Ref.~\cite{Truoel1974} 
and with the theoretical prediction in Ref.~\cite{Phillips1974}.
That means that the non-breakup and breakup channels 
in radiative pion capture in $^3$He are equally important. 

Comparing the total rates for the 
$\pi^- + {^3{\rm H}}  \rightarrow \gamma + n + n + n$
reaction one might draw a false conclusion that 
FSI and 3N force effects are very small. From Fig.~\ref{brdynamics2}
it is, however, clear that the agreement between the plane wave 
and full results for the total rates 
is rather accidental, since the differential rates are quite different.
Contrary to the three-body breakup of $^3$He, FSI effects enhance 
the plane wave result. The 3N force reduces the rate by
approximately 10~\%.
Our result for the total radiative capture rate 
is by about 70~\% larger than the prediction from Ref.~\cite{Phillips1975}.
 
In view of the fact that the theoretical results obtained before 
were based on quite 
different approaches, in many cases the agreement with earlier 
theoretical predictions is satisfactory. In particular, we obtained similar 
shapes of the photon energy spectra for all studied reactions. 
Our predictions about the role of the final state 
interactions based 
on the realistic semi-phenomenological nuclear forces 
are fully converged with respect to the number of partial wave states.
That means that, in contrast to Phillips and Roig~\cite{Phillips1974,Miller1980},
we are ready to calculate not only capture rates but also any polarization observables. 
Our much more advanced model confirms qualitatively 
the $S$-wave based results for capture rates from Refs.~\cite{Phillips1974,Miller1980}.
Thus it will be very interesting to 
compare results of modern calculations performed with improved transition 
operator and consistent nuclear forces.

\begin{table}[ht]
\caption{The total radiative capture rate $\Gamma_{nn}$ in $10^{15}$~1/s 
for the $ \pi^- + {^2{\rm H}} \rightarrow \gamma + n + n $ 
reaction calculated with the AV18~\cite{av18} NN potential
and the non-relativistic single nucleon transition operator~(\ref{jA1}).
Plane wave impulse approximation based results (PW)
and the predictions including nn FSI
(Full) are displayed together with earlier theoretical predictions.
\label{tab1}}
\begin{center}
\begin{tabular}{lcc}
\hline
PW    &   0.318              &      \\
Full  &   0.328              &      \\
earlier theoretical predictions:  &     &              \\
Ref.~\cite{Reitan1966} (1966)  &  0.332  & 0.4 (corrected in Ref.~\cite{Sotona1976}) \\
Ref.~\cite{Sotona1976} (1976)  &  0.375  & (based on pion photoproduction data) \\
Ref.~\cite{Sotona1976} (1976)  &  0.383  & (based on soft-pion limit) \\
Ref.~\cite{Gibbs1977}  (1977)  &  0.420 $\pm$ 0.05  &  \\[4pt]
\hline
\end{tabular}
\end{center}
\end{table}

\begin{table}[ht] 
\caption{Rates $\Gamma$ in $10^{15}$~1/s 
for radiative pion capture in $^3$He and $^3$H
calculated with the AV18~\cite{av18} NN potential
and the non-relativistic single nucleon transition operator~(\ref{jA1}).
Results obtained using the plane wave impulse approximation (PW 2NF),
with consistent treatment of the initial and final nuclear states
based on 2N forces only (Full 2NF) and,  
additionally, employing the Urbana~IX~\cite{urbana} 3N force (Full 2NF+3NF)
are presented.  Earlier theoretical predictions are also displayed.
\label{tab2}}
\begin{center}
\begin{tabular}{lcc}
\hline
$\pi^- + {^3{\rm He}} \rightarrow \gamma + {^3{\rm H}}$  & $\Gamma_{^3{\rm H}}$ & \\
Full 2NF     &   2.059              &      \\
Full 2NF+3NF &   2.132              &      \\
earlier theoretical predictions:  &     &              \\
Ref.~\cite{Fuji1962}        (1962) &  8.32  & 4.28 (corrected in Ref.~\cite{Truoel1974}) \\
Ref.~\cite{Divakaran1965}   (1965) &  0.97  & 3.88 (corrected in Ref.~\cite{Truoel1974}) \\
Ref.~\cite{Griffiths1968}   (1968) &  2.32  &   \\
Ref.~\cite{Pascual1970}     (1970) &  3.37  & 2.25 (corrected in Ref.~\cite{Truoel1974}) \\
Ref.~\cite{Truoel1974}      (1974) &  3.60  &    \\
Ref.~\cite{Phillips1974}    (1974) &  3.1--3.7  &  \\
Ref.~\cite{Gibbs1978}       (1978) &    3.30  &  \\[4pt]
$\pi^- + {^3{\rm He}} \rightarrow \gamma + n + d$ & $\Gamma_{nd}$  & \\
PW 2NF       &   5.201              &      \\
Full 2NF     &   2.013              &      \\
Full 2NF+3NF &   1.840              &      \\
$\pi^- + {^3{\rm He}} \rightarrow \gamma + n + n + p$ & $\Gamma_{nnp}$   & \\
PW 2NF       &   3.816              &      \\
Full 2NF     &   0.659              &      \\
Full 2NF+3NF &   0.615              &      \\[4pt]
$ \left( \Gamma_{nd} + \Gamma_{nnp} \right)/ \Gamma_{^3{\rm H}} $  & & \\
Full 2NF     &   1.30              &      \\
Full 2NF+3NF &   1.15              &      \\[4pt]
earlier theoretical predictions:  &     &              \\
Ref.~\cite{Phillips1974} (1974) &  0.84--1.27  &  \\[4pt]
experimental data:  &     &              \\
Ref.~\cite{Truoel1974}   (1974) &  1.12 $\pm$ 0.05  &   \\[4pt]
$\pi^- + {^3{\rm H}}  \rightarrow \gamma + n + n + n$  & $\Gamma_{nnn}$   & \\
PW 2NF       &   0.117              &      \\
Full 2NF     &   0.141              &      \\
Full 2NF+3NF &   0.128              &      \\
earlier theoretical predictions:  &     &              \\
Ref.~\cite{Phillips1975} (1975)  &  0.07  &  \\[4pt]
\hline
\end{tabular}
\end{center}
\end{table}

\section{Summary and conclusions}
\label{section8}

Recent theoretical work by G{\aa}rdestig and Phillips \cite{Gardestig2006PRC2,Gardestig2006PRL}
shows that radiative pion capture is not only interesting by itself 
but also correlated with a variety of other processes when studied 
within chiral effective field theory. This is very important 
because the other reactions (like muon capture) or neutrino induced 
processes are much harder to measure and the information 
from the radiative capture on light nuclei could significantly improve 
our understanding of the other processes. 
Thus a uniform framework 
for the calculations of various electromagnetic and weak reactions
on the single nucleon, deuteron, $^3$He, $^3$H and other light nuclei 
should be formulated and applied. This framework 
would comprise consistent two-nucleon and more-nucleon forces
as well as transition operators (``currents'') with one-body 
and many-body parts. 
Results of fully converged calculations should be ultimately
compared with precise experimental data, to yield a broad and complete picture 
of these reactions in few-nucleon systems.

In the present paper, we studied the
$\pi^- + {^2{\rm H}} \rightarrow \gamma + n + n$,
$\pi^- + {^3{\rm He}} \rightarrow \gamma + {^3{\rm H}}$,
$\pi^- + {^3{\rm He}} \rightarrow \gamma + n + d$,
$\pi^- + {^3{\rm He}} \rightarrow \gamma + n + n + p$
and
$\pi^- + {^3{\rm H}} \rightarrow \gamma + n + n + n$
reactions using traditional nuclear forces (the AV18 NN 
potential and the Urbana~IX 3N force) and a simple 
single-nucleon transition operator. These calculations, like 
our studies of muon capture \cite{PRC90.024001,PRC94.034002} or very recent investigations
of some neutrino induced reactions~\cite{Golak2018},
are ready to be systematically improved to encompass more complicated dynamical input.  
Many aspects of the performed calculations, like 
the role of the relativistic kinematics,
the efficient methods of partial wave decomposition or
the convergence of our results with respect to the number
of partial wave states, have been already established
and the predictions presented here can serve 
as an important benchmark.

Our calculations already provide first realistic predictions 
for the differential 
$d\Gamma_{^3{\rm H}} /dE_{\gamma}$,
$d\Gamma_{nd} /dE_{\gamma}$,
$d\Gamma_{nnp} /dE_{\gamma}$ 
and
$d\Gamma_{nnn} /dE_{\gamma}$ 
capture rates as well as for the corresponding total radiative capture rates
$\Gamma_{^3{\rm H}}$,
$\Gamma_{nd}$,
$\Gamma_{nnp}$ 
and
$\Gamma_{nnn}$.
The formalism used in the present paper
will be in the future extended to study other pion capture reactions,
including non-radiative and double radiative pion capture.

\acknowledgments
We would like to thank Peter Tru\"ol for drawing our attention 
to radiative pion capture and for providing us with crucial references.
This work is a part of the LENPIC project and was supported
by the Polish National Science Centre under Grants
No. 2016/22/M/ST2/00173 and 2016/21/D/ST2/01120
and by DFG and NSFC through funds provided 
to the Sino-German CRC 110 “Symmetries and the Emergence of Structure in
QCD” (NSFC (11621131001), DFG (TRR110)).
The numerical calculations were
partially performed on the supercomputer cluster of the JSC, J\"ulich, Germany.


\begin{thebibliography}{99}

\bibitem{Panofsky1951} 
W.~K.~H.~Panofsky, R. Lee Aamodt, and J. Hadley, 
Phys. Rev. {\bf 81}, 565 (1951).

\bibitem{Baer1977}
H.~W.~Baer, K.~M.~Crowe, and P.~Tru\"ol,
Adv. Nucl. Phys. {\bf 9}, 177 (1977).

\bibitem{Renker1978}
D. Renker {\em et al.}, 
Phys. Rev. Lett. {\bf 41}, 1279 (1978).

\bibitem{Reynaud1981}
G. W. Reynaud and F. Tabakin,
Phys. Rev. C{\bf 23}, 2652 (1981).

\bibitem{Martoff1983}
C. J. Martoff {\em et al.},
Phys. Rev. C{\bf 27}, 1621 (1983).

\bibitem{Baer1983} 
H. W. Baer {\em et al.},
Phys. Rev. C{\bf 28}, 761 (1983).

\bibitem{Roig1985}
F. Roig and J. Navarro, 
Nucl. Phys. A{\bf 440}, 659 (1985).

\bibitem{Singham1986}
M. K. Singham and F. Tabakin,
Phys. Rev. C{\bf 34}, 637 (1986).

\bibitem{Krivine1988}
H. Krivine, E. Lipparini, J. Navarro and F. Roig, 
Nucl. Phys. A{\bf 481}, 781 (1988).

\bibitem{Navarro1989}
J. Navarro and  F. Roig
Phys. Rev. C{\bf 39}, 302 (1989).

\bibitem{Raywood1997}
K. J. Raywood {\em et al.}, 
Phys. Rev. C{\bf 55}, 2492 (1997).

\bibitem{Amaro1997}
J. E. Amaro, A. M. Lallena and J. Nieves,
Nucl. Phys. A{\bf 623}, 529 (1997).

\bibitem{Watson1951} K. M. Watson and R. N. Stuart, 
Phys. Rev. {\bf 82}, 738 (1951).

\bibitem{McVoy1961} K. McVoy, 
Phys. Rev. {\bf 121}, 1401 (1961).

\bibitem{Bander1964}
M. Bander,
Phys. Rev. {\bf 134}, B1052 (1964).

\bibitem{Gibbs1975} W. R. Gibbs, B. F. Gibson, and Q. J. Stephenson, Jr.,
Phys. Rev. C{\bf 11}, 90 (1975); Erratum: [Phys. Rev. C{\bf 12}, 2130 (1975)].

\bibitem{Gibbs1977} W. R. Gibbs, B. F. Gibson, and Q. J. Stephenson, Jr.,
Phys. Rev. C{\bf 16}, 327 (1977); Erratum: [Phys. Rev. C{\bf 17}, 856 (1978)].

\bibitem{deTeramond1977} 
G. F. de T\'eramond,
Phys. Rev. C{\bf 16}, 1976 (1977).

\bibitem{deTeramond1987} 
G. F. de T\'eramond,
Phys. Rev. C{\bf 36}, 691 (1987).

\bibitem{Gardestig2006PRC1}
A. G{\aa}rdestig and D. R. Phillips, 
Phys. Rev. C{\bf 73}, 014002 (2006).

\bibitem{Gardestig2006PRL}
A. G{\aa}rdestig and D. R. Phillips, 
Phys. Rev. Lett. {\bf 96}, 232301 (2006).

\bibitem{Phillips1954} 
R. H. Phillips and K. M. Crowe, 
Phys. Rev. {\bf 96}, 484 (1954).

\bibitem{Ryan1964} 
J. W. Ryan, 
Phys. Rev. Lett. {\bf 12}, 564 (1964).

\bibitem{Haddock1965} 
P. Haddock, R. M. Salter, Jr. , M. Zeller, J. B. Czirr, and D.  R. Nygren,
Phys. Rev. Lett. {\bf 14}, 318 (1965).

\bibitem{Nicholson1968}
J. P. Nicholson, P. G. Butler, N. Cohen, and A. N. James,
Phys. Lett. B{\bf 27}, 452 (1968).

\bibitem{Salter1975} 
R. M. Salter, Jr. , R. P. Haddock, M. Zeller, D. R. Nygren, and J. B. Czirr,
Nucl. Phys. A{\bf 254}, 241 (1975).

\bibitem{Gabioud1979}
B. Gabioud {\em et al.},
Phys. Rev. Lett. {\bf 42}, 1508 (1979).

\bibitem{Gabioud1981}
B. Gabioud {\em et al.}, 
Phys. Lett. B{\bf 103}, 9 (1981).

\bibitem{Gabioud1984}
B. Gabioud {\em et al.},
Nucl. Phys. A{\bf 420}, 496 (1984).

\bibitem{Schori1987}
O. Schori {\em et al.},
Phys. Rev. C{\bf 35}, 2252 (1987).

\bibitem{Howell1998}
C.~R.~Howell {\em et al.},
Phys. Lett. B{\bf 444}, 252 (1998).

\bibitem{Chen2008} 
Q. Chen {\em et al.},
Phys. Rev. C{\bf 77}, 054002 (2008).

\bibitem{Slaus1989}
I. Slaus, Y. Akaishi, and H. Tanaka, Phys. Rept. {\bf 173}, 257 (1989).

\bibitem{Gardestig2009} 
A. G{\aa}rdestig,
J. Phys. G: Nucl. Phys. {\bf 36}, 053001 (2009).

\bibitem{Huhn2000}
V. Huhn, L. W\"atzold, Ch. Weber, A. Siepe, W. von Witsch, H. Wita{\l}a, and W. Gl\"ockle,
Phys. Rev. Lett. {\bf 85}, 1190 (2000). 

\bibitem{GonzalezTrotter2006}
D. E. Gonzalez Trotter {\em et al.},
Phys. Rev. C{\bf 73}, 034001 (2006).

\bibitem{Gibbs1977.1} W. R. Gibbs, B. F. Gibson, and Q. J. Stephenson, Jr.,
Phys. Rev. C{\bf 16}, 322 (1977).

\bibitem{Reid1968}
R. V. Reid,
Ann. Phys. {\bf 50}, 411 (1968).

\bibitem{Muskhelishvili}
N. I. Muskhelishvili, {\it Singular Integral Equations},
(Noordhoff, Groningen, The Netherlands, 1953).

\bibitem{Omnes}
R. Omn\`es, Nuovo Cimento {\bf 8}, 316 (1958); M. Jacob,
G. Mahoux and R. Omn\`es, ibid. {\bf 23}, 838 (1962).

\bibitem{deTeramond1980} 
G. F. de T\'eramond, J. P\'aez, and C. W. Soto Vargas,
Phys. Rev. C{\bf 21}, 2542 (1980).

\bibitem{Bernard1996.1}
V. Bernard, N. Kaiser, U.-G. Mei{\ss}ner,
Z. Phys. C{\bf 70}, 483 (1996).

\bibitem{Bernard1996.2}
V.~Bernard, N.~Kaiser, and U.-G.~Mei{\ss}ner, 
Phys. Lett. B{\bf 383}, 116, (1996).

\bibitem{Fearing2000}
H.~W.~Fearing, T.~R.~Hemmert, R. Lewis, and Ch. Unkmeir,
Phys. Rev. C{\bf 62}, 054006 (2000).

\bibitem{Gardestig2006PRC2}
A.~G{\aa}rdestig, 
Phys. Rev. C{\bf 74}, 017001 (2006).

\bibitem{Messiah1952}
A. M. L. Messiah,
Phys. Rev. {\bf 87}, 639 (1952).

\bibitem{KR1954}
N.~M.~Kroll and M.~A.~Ruderman, Phys. Rev. {\bf 93}, 233 (1954).

\bibitem{Divakaran1965}
P. P. Divakaran, 
Phys. Rev. {\bf 139}, 3887 (1965).

\bibitem{Ericson1967}
M. Ericson, A. Figureau, 
Nucl. Phys. B{\bf 3}, 609 (1967).

\bibitem{Ericson1969}
M. Ericson, A. Figureau, 
Nucl. Phys. B{\bf 11}, 621 (1969).

\bibitem{Griffiths1968}
D. Griffiths and C. Kim,
Phys. Rev. {\bf 173}, 1584 (1968).

\bibitem{Pascual1970}
P. Pascual and A. Fujii,
Nuovo Cimento {\bf 65}, 411 (1970).   

\bibitem{Zaimidoroga65}
O. A. Zaimidoroga {\em et al.},
JETP (Sov. Phys.) {\bf 21}, 848 (1965).

\bibitem{Zaimidoroga67}
O. A. Zaimidoroga {\em et al.},
JETP (Sov. Phys.) {\bf 24}, 1111 (1967).

\bibitem{Phillips1974}
A. C. Phillips and F. Roig,
Nucl. Phys. A{\bf 234}, 378 (1974).

\bibitem{Gibbs1978}
W. R. Gibbs, B. F. Gibson, and Q. J. Stephenson, Jr.,
Phys. Rev. C{\bf 18}, 1761 (1978).

\bibitem{Truoel1974}
P. Tru\"ol, H. W. Baer, J. A. Bistirlich, K. M. Crowe, N. de Botton and J. A. Helland,
Phys. Rev. Lett. {\bf 32}, 1268 (1974).

\bibitem{Fuji1962}
A. Fuji and D. J. Hall,
Nucl. Phys. {\bf 32}, 102 (1962).

\bibitem{Delorme1966}
J. Delorme and T. E. O. Ericson,
Phys. Lett. {\bf 21}, 98 (1966).

\bibitem{Amado1963}
R. D. Amado, 
Phys. Rev. {\bf 132}, 485 (1963).

\bibitem{Lovelace1964}
C. Lovelace,
Phys. Rev. {\bf 135}, B1225 (1964).

\bibitem{Phillips1975}
A. C. Phillips and F. Roig, in {\it High Energy Physics
and Nuclear Structure}, edited by D. E. Nagle {\em et al.},
AIP Conference Proceedings No. 26 (American Institute
of Physics, New York, 1975).

\bibitem{Bistirlich1976}
J. A. Bistirlich {\em et al.},
Phys. Rev. Lett. {\bf 36}, 942 (1976).

\bibitem{Miller1980}
J. P. Miller {\em et al.},
Nucl. Phys. A{\bf 343}, 341 (1980).

\bibitem{PRC90.024001}
J.~Golak, R.~Skibi\'nski, H.~Wita{\l}a,
K.~Topolnicki, A.~E.~Elmeshneb, H.~Kamada, A.~Nogga, and L.~E.~Marcucci,
Phys. Rev. C{\bf 90}, 024001 (2014).

\bibitem{PRC94.034002}
J.~Golak, R.~Skibi\'nski, H.~Wita{\l}a,
K.~Topolnicki, H.~Kamada, A.~Nogga, and L.~E.~Marcucci,
Phys. Rev. C{\bf 94}, 034002 (2016).

\bibitem{prc83.014002} 
L.E.\ Marcucci, M.\ Piarulli, M.\ Viviani, L.\ Girlanda, A.\ Kievsky, 
S.\ Rosati, and R.\ Schiavilla,
Phys.\ Rev.\ C {\bf 83}, 014002 (2011).

\bibitem{Kie08}
A.\ Kievsky, S.\ Rosati, M.\ Viviani, L.E.\ Marcucci, and L.\ Girlanda,
J.\ Phys.\ G {\bf 35}, 063101 (2008).

\bibitem{PRC86.035503}
G.~Shen, L.~E.~Marcucci, J.~Carlson, S.~Gandolfi, and R.~Schiavilla,
Phys. Rev. C{\bf 86}, 035503 (2012).
 
\bibitem{av18} R.B. Wiringa, V.G.J. Stoks, and R. Schiavilla,
Phys. Rev. C {\bf 51}, 38 (1995).

\bibitem{urbana} B.S. Pudliner, V.R. Pandharipande, J. Carlson, Steven C. Pieper, and R.B. Wiringa,
Phys. Rev. C {\bf 56}, 1720 (1997).

\bibitem{edis3d} K. Topolnicki, J. Golak, R. Skibi\'nski, A.E. Elmeshneb, W. Gl\"ockle, A. Nogga, and H. Kamada,
Few-Body Syst. {\bf 54}, 2223 (2013).

\bibitem{physrep} J. Golak, R. Skibi\'nski, H. Wita{\l}a, W. Gl\"ockle, A.  Nogga, and H. Kamada,
Phys. Rept. {\bf 415}, 89 (2005).

\bibitem{romek2} R. Skibi\'nski, J. Golak, H. Wita{\l}a, W. Gl\"ockle, and A. Nogga,
Eur. Phys. J. A {\bf 24}, 11 (2005).

\bibitem{Reitan1966}
A. Reitan,
Nucl. Phys. {\bf 87}, 232 (1966).

\bibitem{Sotona1976}
M. Sotona and E. Truhlik, 
Nucl. Phys, A{\bf 262}, 400 (1976).

\bibitem{Golak2018}
J. Golak, R. Skibi\'nski, K. Topolnicki, H. Wita{\l}a, A. Grassi, H. Kamada, L. E. Marcucci,
Phys. Rev. C{\bf 98}, 015501 (2018).

\end{thebibliography}
\end{document}